\documentclass[12pt]{article}
\usepackage{amsmath}
\usepackage{amssymb}
\usepackage{graphicx}
\usepackage{ifthen}
\usepackage{verbatim}

\newlength{\captionwidth}
\newsavebox{\tempbox}
\newcommand{\mycaption}[2]{%
\par\vspace{10pt}\sbox{\tempbox}{Figure #1: #2}%
\ifthenelse{\lengthtest{\wd\tempbox>\captionwidth}}%
{\sbox{\tempbox}{Figure.#1:\ }%
\addtolength{\captionwidth}{-\wd\tempbox}%
\mbox{Figure #1:\ }\parbox[t]{\captionwidth}{\small\textit{#2}}}%
{Figure #1: {\small\textit{#2}}}}%

\numberwithin{equation}{section}

\allowdisplaybreaks
\begin{document}
\thispagestyle{empty}
\begin{flushright}
October  2007 \\
February 2008 (revised)\\
April 2008 (revised)
\end{flushright}
\bigskip
\bigskip
\begin{center}
{\Large 
\textbf{Melting Crystal, Quantum Torus and Toda Hierarchy}\\
} 
\end{center}
\bigskip
\bigskip
\renewcommand{\thefootnote}{\fnsymbol{footnote}}
\begin{center}
Toshio Nakatsu
\footnote{E-mail: \texttt{nakatsu@phys.sci.osaka-u.ac.jp}}$^1$
and 
Kanehisa Takasaki
\footnote{E-mail: \texttt{takasaki@math.h.kyoto-u.ac.jp}}$^2$\\
\bigskip
{\small
\textit{$^1$Department of Physics, Graduate School of Science,
Osaka University,\\
Toyonaka, Osaka 560-0043, Japan}}\\
{\small
\textit{$^2$Graduate School of Human and Environmental Studies, 
Kyoto University,\\ 
Yoshida, Sakyou, Kyoto 606-8501, Japan}}
\end{center}
\bigskip
\bigskip
\renewcommand{\thefootnote}{\arabic{footnote}}
\begin{abstract}
Searching for the integrable structures 
of supersymmetric gauge theories and topological strings,  
we study melting crystal, 
which is known as random plane partition, 
from the viewpoint of integrable systems. 
We show that 
a series of partition functions of 
melting crystals gives rise to a tau function of 
the one-dimensional Toda hierarchy, 
where the models are defined by 
adding suitable potentials, 
endowed with a series of coupling constants, 
to the standard statistical weight.  
These potentials can be converted to 
a commutative sub-algebra of quantum torus Lie algebra. 
This perspective reveals a remarkable connection  
between random plane partition and quantum torus Lie algebra, 
and substantially enables to prove the statement. 
Based on the result, 
we briefly argue the integrable structures of 
five-dimensional $\mathcal{N}=1$ supersymmetric gauge theories 
and 
$A$-model topological strings. 
The aforementioned potentials correspond to 
gauge theory observables analogous to the Wilson loops, 
and thereby the partition functions are translated 
in the gauge theory 
to generating functions of their correlators. 
In topological strings, 
we particularly comment on a possibility of topology change 
caused by condensation of these observables, 
giving a simple example.

\end{abstract}

\setcounter{footnote}{0}
\newpage


\section{Introduction}

An unanticipated but very exciting connection between 
the statistical mechanical problem of melting crystal, 
known as \textit{random plane partition}, 
and 
$A$-model topological strings 
has been revealed \cite{Crystal}, 
based on the topological vertex \cite{Iqbal, Aganagic-Klemm-Marino-Vafa}. 
The topological vertex is a diagrammatical method which enables 
to compute all genus topological $A$-model string amplitudes 
for a certain class of local geometries.

\begin{figure}[h]
\begin{minipage}{80mm}
\hspace{15mm}
\includegraphics[scale=.3]{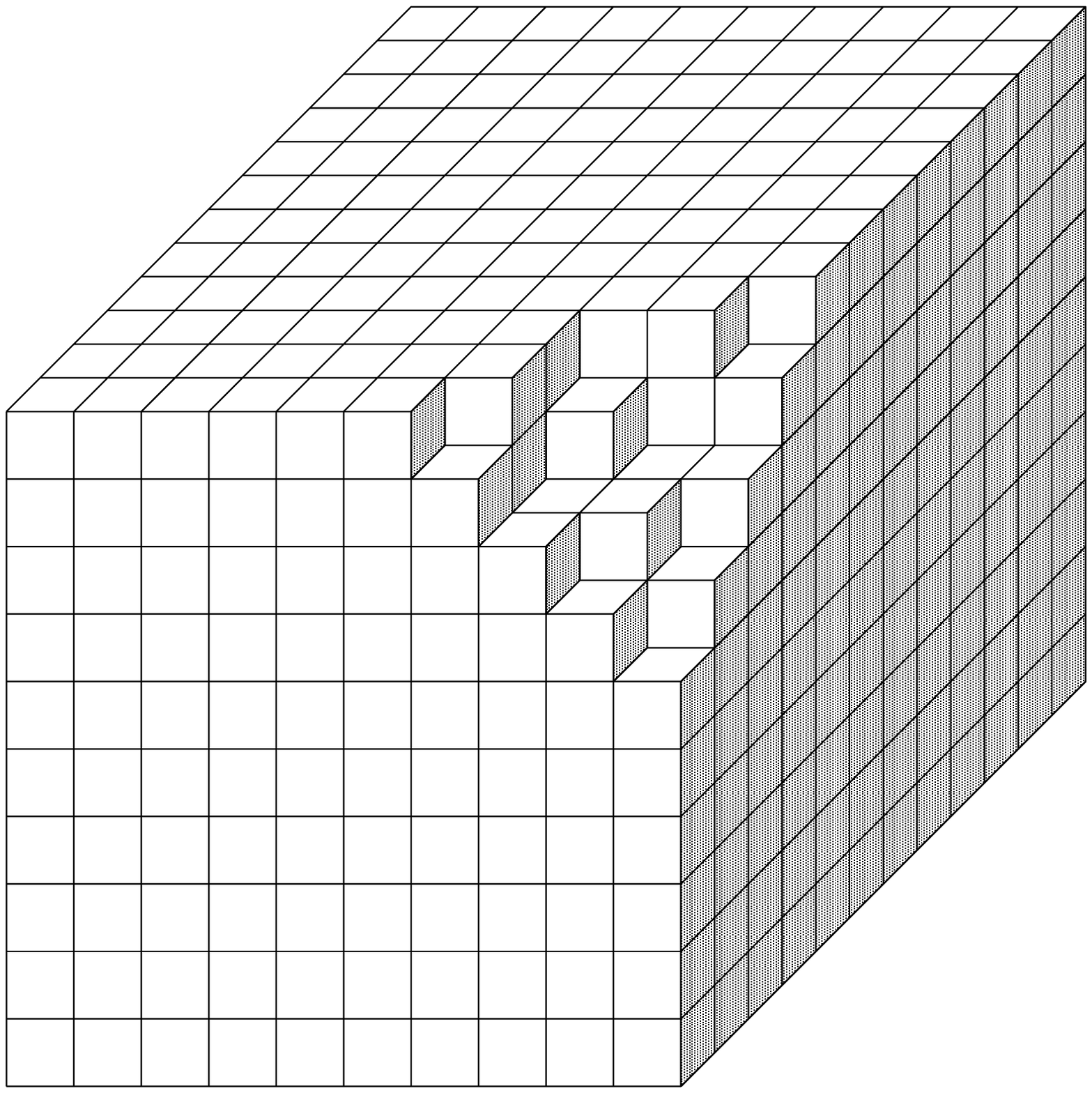}
\end{minipage}
\begin{minipage}{80mm}
\includegraphics[scale=0.5]{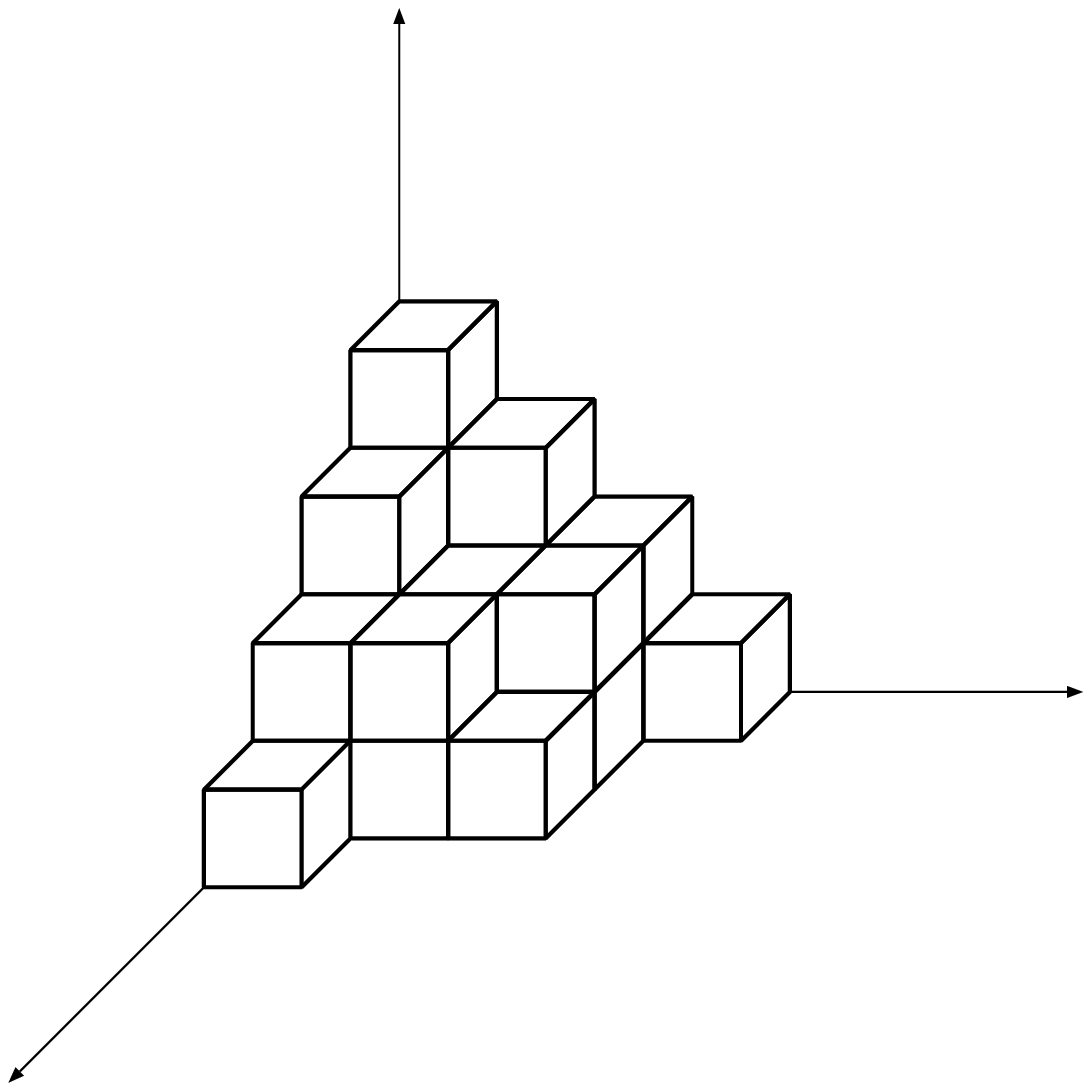}
\end{minipage}
\caption{The corner of the melting crystal and 
the corresponding plane partition 
(the three-dimensional Young diagram).}
\label{crystal and plane partition}
\end{figure}

We can image the positive octant 
$\mathbb{Z}_{\geq 0}^3 \subset \mathbb{R}^3$
occupied by unit cubes
as the neighborhood of a corner of the crystal
by putting unit cubes on the lattice points in the octant.
The frozen crystal occupies the positive octant 
$\mathbb{Z}_{\geq 0}^3$.
As the crystal melts, 
we remove atoms from the corner.
We identify the configuration of crystal melting
as the configuration of plane partition 
or the three-dimensional Young diagrams, 
as depicted in Figure \ref{crystal and plane partition}. 
Removing each atom contributes the factor $q=e^{-\frac{\mu}{T}}$
to the Boltzmann weight of the configuration,
where $T$ is the temperature and $\mu$ is the chemical potential.
Heating up the crystal
leads to melting of it.
%
%

Random plane partition also has a significant relation with 
five-dimensional $\mathcal{N}=1$ supersymmetric gauge theories. 
Nekrasov's formula \cite{Nekrasov, Nekrasov-Okounkov} 
for five-dimensional $\mathcal{N}=1$ supersymmetric 
$SU(N)$ Yang-Mills theory 
can be retrieved from the partition function 
of random plane partition \cite{MNTT1}, 
where the model is interpreted as a $q$-deformed random partition. 
All genus topological $A$-model string amplitude 
for the local $SU(N)$ geometry is evaluated by the topological vertex 
and reproduces Nekrasov's formula for $\mathcal{N}=2$ $SU(N)$ gauge theory 
\cite{Iqbal-Kashani-Poor, Eguchi-Kanno}, 
as predicted in the geometric engineering.

It is shown in \cite{Nekrasov-Okounkov} that
the Seiberg-Witten solutions \cite{Seiberg-Witten} 
of four-dimensional $\mathcal{N}=2$ supersymmetric gauge theories  
emerge through \textit{random partition}, 
where 
Nekrasov's functions for four-dimensional $\mathcal{N}=2$ 
supersymmetric gauge theories are 
understood as the partition functions of random partition. 
The integrable structure of random partition is elucidated in 
\cite{Nekrasov-Marshakov}, and 
thereby the integrability of correlation functions 
among certain observables in four-dimensional 
$\mathcal{N}=2$ supersymmetric gauge theories is explained.

Motivated by the these results, 
we study in this article 
the integrable structure of random plane partition 
in order to search for the integrable structures of 
supersymmetric gauge theories and topological strings.

A partition $\lambda=(\lambda_1,\lambda_2,\cdots)$ 
is a sequence of non-negative integers 
satisfying $\lambda_{i} \geq \lambda_{i+1}$ for all $i\geq 1$.
Partitions are identified with the Young diagrams. 
The size is defined by $|\lambda|=\sum_{i \geq 1}\lambda_i$, 
which is the total number of boxes of the diagram.
A plane partition $\pi$ is an array of 
non-negative integers 
\begin{eqnarray}
\begin{array}{cccc}
\pi_{11} & \pi_{12} & \pi_{13} & \cdots \\
\pi_{21} & \pi_{22} & \pi_{23} & \cdots \\
\pi_{31} & \pi_{32} & \pi_{33} & \cdots \\
\vdots & \vdots & \vdots & ~
\end{array}
\label{pi}
\end{eqnarray}
satisfying 
$\pi_{ij}\geq \pi_{i+1 j}$ and $\pi_{ij}\geq \pi_{i j+1}$ 
for all $i,j \geq 1$. 
Plane partitions are identified 
with the three-dimensional Young diagrams. 
The three-dimensional diagram $\pi$ 
is a set of unit cubes such that $\pi_{ij}$ cubes 
are stacked vertically on each $(i,j)$-element of $\pi$. 
The size is defined by 
$|\pi|=\sum_{i,j \geq 1}\pi_{ij}$,  
which is the total number of cubes of the diagram. 
Diagonal slices of a plane partition $\pi$ become partitions, 
as depicted in Figure 2.     
Let $\pi(m)$ denote the partition 
along the $m$-th diagonal slice, 
where $m \in \mathbb{Z}$. 
In particular, 
$\pi(0)=(\pi_{11},\pi_{22},\cdots)$ 
is the main diagonal partition.  
This series of partitions satisfies the condition
\begin{eqnarray}
\cdots \prec \pi(-2) \prec \pi(-1) \prec 
\pi(0) \succ \pi(1) \succ \pi(2) \succ \cdots,
\label{interlace relations}
\end{eqnarray}
where $\mu \succ \nu$ means the interlace relation 
between two partitions $\mu,\nu$.  
\begin{eqnarray}
\mu \succ \nu ~~~
\Longleftrightarrow ~~~
\mu_1 \geq \nu_1 \geq \mu_2 \geq \nu_2 
\geq \mu_3 \geq \cdots.
\end{eqnarray} 
%
%
\begin{figure}[h]
\begin{center}
\includegraphics[scale=0.55]{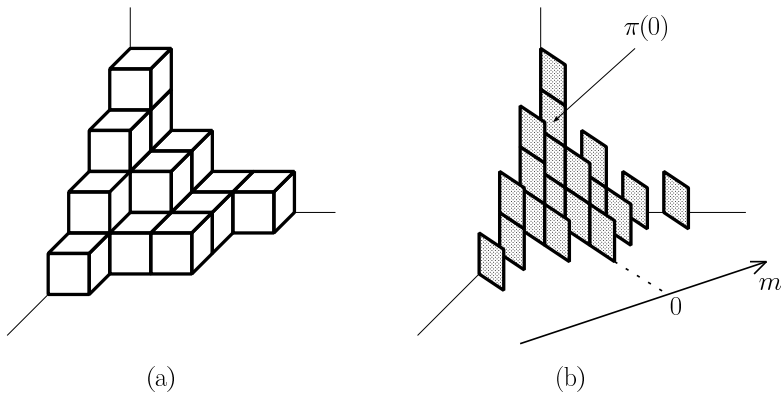}
\caption{Plane partition 
(The three-dimensional Young diagram) (a) 
and the corresponding sequence of partitions 
(the two-dimensional Young diagrams) (b).}
\end{center}
\label{three-dimensional Young diagram}
\end{figure}
%
%

A statistical model of plane partitions is introduced 
by the following partition function.
\begin{eqnarray}
Z 
&\equiv& 
\sum_{\pi}q^{|\pi|}\,,
\label{Z}
\end{eqnarray}
where the sum is over all plane partitions.
The parameter $q$ is indeterminate satisfying $0<q<1$ 
and play a role of chemical potential
(the energy of the removal of an atom from the crystal).
Summing over plane partitions, we can obtain
\begin{eqnarray}
\sum_{\pi}~q^{|\pi|}=\prod_{n=1}^{\infty}~\frac{1}{(1-q^n)^n}.
\label{McMahon}
\end{eqnarray}
The partition function of the model
is the generating function of plane partitions,
known as the McMahon function.

\subsection{The models}
We introduce a series of potentials for partitions 
and put the main diagonal partition of $\pi$
in these potentials. 
For the later convenience, 
we introduce them as functions on charged partitions,  
which are partitions paired with integers.   
Let $\Phi_k$ ($k=1,2,\cdots$) be 
the following functions on charged partitions.@ 
\begin{eqnarray}
\Phi_k(\lambda,p)\,
=\,
\sum_{i=1}^{\infty}q^{k(p+\lambda_i-i+1)}
-\sum_{i=1}^{\infty}q^{k(-i+1)}\,.  
\label{Phi_k}
\end{eqnarray}
where $(\lambda,p)$ denotes a charged partition. 
Actually, the right hand side of this formula 
becomes a finite sum by cancellation of terms 
between the two sums.  More precisely, 
\begin{eqnarray}
  \Phi_k(\lambda,p)
  =\sum_{i=1}^\infty 
  (q^{k(p+\lambda_i-i+1)} - q^{k(p-i+1)}) 
  +q^k\frac{1-q^{pk}}{1-q^k}\,.
\end{eqnarray}
With each fixed value of $p$,  
these provide a series of potentials for partitions. 
These functions have been exploited in 
\cite{Nekrasov-Marshakov} 
from the four-dimensional gauge theory viewpoint, 
with $q$ or $q^k$ being replaced 
by a generating spectral parameter. 
Introducing the coupling constants $t=(t_1,t_2,\cdots )$,  
we write their combination as 
\begin{eqnarray}
\Phi_{(t;\,p)}(\lambda)
=
\sum_{k=1}^{\infty}
t_k \Phi_k(\lambda, p)\,.  
\label{Phi_(t,p)}
\end{eqnarray}
The partition function of 
random plane partition 
whose main diagonal partition is 
in the potential (\ref{Phi_(t,p)}) 
is defined by 
\begin{eqnarray}
Z_p(t)\equiv
\sum_{\pi}
q^{|\pi|}
e^{\Phi_{(t;\,p)}(\pi(0))}
\,.
\label{Z_p(t)}
\end{eqnarray}

The model has an interpretation as a $q$-deformed random partition.  
To see this, 
note that, by virtue of 
the interlacing relations (\ref{interlace relations}), 
the two series of partitions 
$\bigl\{\pi(m)\bigr\}_{m=0}^\infty$ 
and 
$\bigl\{\pi(-m)\bigr\}_{m=0}^\infty$ 
represent a pair $T,T'$ of semi-standard Young tableaux 
of shape $\pi(0)$, 
in which the part of the $m$-th skew 
Young diagrams $\pi(\pm m)/\pi(\pm(m+1))$ is filled with $m+1$. 
The partition function can be thereby reorganized to 
a sum over the Young diagram $\lambda = \pi(0)$ and 
the pair $T,T'$ of semi-standard Young tableaux 
of shape $\lambda$ as 
\begin{eqnarray}
  Z_p(t) 
= \sum_{\lambda} 
   \sum_{T,T'\,:\,\mathrm{shape}\,\lambda}
    q^T q^{T'}e^{\Phi_{(t;\,p)}(\lambda)}, 
\end{eqnarray}
where 
$\displaystyle{
q^T = \prod_{m=0}^\infty q^{(m+\frac{1}{2})(|\pi(m)|-|\pi(m+1)|)}
}$ 
and 
$\displaystyle{
q^{T'} = \prod_{m=0}^\infty q^{(m+\frac{1}{2})(|\pi(-m)|-|\pi(-m-1)|)}
}$. 
The partial sum over the semi-standard tableaux 
gives the Schur function $s_\lambda(q^{\rho}) 
= s_\lambda(x_1,x_2,\cdots)$ specialized to 
$x_i = q^{i-\frac{1}{2}}$: 
\begin{eqnarray}
  \sum_{T:\,\mathrm{shape}\,\lambda}q^T 
= \sum_{T':\,\mathrm{shape}\,\lambda}q^{T'}
= s_\lambda(q^\rho). 
\end{eqnarray}
Therefore the partition function can be eventually 
expressed as 
\begin{eqnarray}
Z_p(t)=
\sum_{\lambda}\,
e^{\Phi_{(t;\,p)}(\lambda)}\,
s_{\lambda}(q^{\rho})^2. 
\label{Z_p(t) by Schur}
\end{eqnarray} 
The representation of $s_{\lambda}(q^{\rho})$ 
in terms of the hook polynomial \cite{Macdonald} 
allows us to write it further as 
\begin{eqnarray}
Z_p(t)=
\sum_{\lambda}\,
e^{\Phi_{(t;\,p)}(\lambda)}\,
\left(
  \frac{q^{\frac{1}{2}|\lambda|+n_{\lambda}}}
     {\prod_{s \in \lambda}(1-q^{h(s)})}
\right)^2
\,, 
\label{Z_p(t) by Hook polynomial}
\end{eqnarray} 
where $h(s)$ denotes the hook length 
of the box $s \in \lambda$.

\subsection{Main result}
In this article, 
we study these models of random plane partition  
from the viewpoint of integrable systems. 
Our main result is that 
the following series of the partition functions 
is a tau function of an integrable hierarchy: 
\begin{eqnarray}
\tau(t;p)\equiv 
q^{\frac{1}{6}p(p+1)(2p+1)}Z_p(t)\,, 
\hspace{8mm}
p \in \mathbb{Z}\,.
\label{tau(t;p)=Z_p}
\end{eqnarray}
More precisely, 
we show in the text that 
$\tau(t;p)$ is a tau function of 
the one-dimensional Toda hierarchy,  
where the coupling constants $t=(t_1,t_2,\cdots)$ 
are interpreted as a single series 
of time variables of the one-dimensional Toda hierarchy.
In particular,  
$\tau(t;p)$ is shown to have a representation   
by two-dimensional free fermions or free bosons as 
\begin{eqnarray}
\tau(t;p)
&=&
e^{\sum_{k=1}^{\infty}\frac{t_kq^k}{1-q^k}}\,\,
\langle p |\,
e^{\frac{1}{2}\sum_{k=1}^{\infty}(-)^kt_kJ_k}
\,\,
{\bf g}_{\star}
\,\, 
e^{\frac{1}{2}\sum_{k=1}^{\infty}(-)^kt_kJ_{-k}}\,\, 
|p \rangle\,. 
\label{Z_p(t) Toda}
\end{eqnarray}
In the right hand side of this formula,  
${\bf g}_{\star}$ is an element of $GL(\infty)$ 
and  $J_{\pm k}$ 
denote the modes of the $U(1)$ current 
associated with the complex fermions. 
Thus $\tau(t;p)$ is a matrix element, 
taken between the Dirac sea of $U(1)$ charge $p$, 
of an element of $GL(\infty)$.  The exponential 
operators on both hand sides of ${\bf g}_{\star}$ 
are generators of the commuting flows of 
the one-dimensional Toda hierarchy.


Once it is known that the partition function relates with the tau function, 
it becomes amenable to obtain an infinite set of 
non-linear differential equations that the partition function obeys. 
This is because tau functions of an integrable hierarchy   
satisfy an infinite set of non-linear differential equations.
These non-linear differential equations are all encoded 
in the bilinear identity. 
It follows from the above that the partition functions satisfy 
the following bilinear identities in the one-dimensional Toda hierarchy. 
\begin{eqnarray}
&&
\oint \frac{dz}{2\pi i} 
z^{p-p'}
e^{\frac{1}{2}\sum_{k \geq 1}(t_k-t'_k)z^k} 
Z_{p}(t-[z^{-1}])Z_{p'}(t'+[z^{-1}])
\nonumber \\
&&
=
q^{(p+p'+1)(p-p'+1)}
\oint \frac{dz}{2\pi i} 
z^{p-p'}
e^{-\frac{1}{2}\sum_{k \geq 1}(t_k-t'_k)z^{-k}} 
Z_{p+1}(t+[z])Z_{p'-1}(t'-[z])\,,  
\label{Toda bilinear identity}
\end{eqnarray}
where $p,p'$ are arbitrary. 
The integral on the left hand side means 
taking the residue at $z = \infty$ and multiply it 
by $-1$; the integral on the right hand side 
is understood to be the residue at $z = 0$. 
We also use notations like 
$t \pm [z]=
(t_1 \pm z, t_2 \pm \frac{z^2}{2}, t_3 \pm \frac{z^3}{3}, \cdots)$.  
Towers of non-linear differential equations 
are obtained from (\ref{Toda bilinear identity}) 
as the coefficients of the Taylor expansions along the diagonal $t=t'$, 
that is, 
the coefficients of the expansions 
of the bilinear identities in the variables $t_k-t'_k$. 
For instance, let $p=p'$. The first equation 
one gets in the tower is nothing but 
the Toda equation in Hirota's bilinear form. 
\begin{eqnarray}
D_{t_1}^2Z_p \cdot Z_p=q^{2p+1}Z_{p+1}Z_{p-1}\,, 
\label{Toda equation in Hirota form}
\end{eqnarray}
where $D$ denotes Hirota's derivative   
that is defined by 
$\displaystyle D_x f(x) \cdot g(x) 
=\lim_{y \rightarrow x}
(\partial_x-\partial_y)f(x)g(y)$.

The partition functions also give rise to 
a solution of the modified KP hierarchy. 
The corresponding bilinear identities are read as 
\begin{eqnarray}
\oint \frac{dz}{2\pi i} 
z^{p-p'}
e^{\sum_{k \geq 1}(t_k-t'_k)z^k} 
Z_{p}(t-[z^{-1}])Z_{p'}(t'+[z^{-1}])
=0\,, 
\label{MKP bilinear identity}
\end{eqnarray}
where $p \geq p'$. 
These identities include in the towers 
the modified KP equations as well as the KP equation. 
For instance, let $p'=p-1$. 
The first equation one gets in this tower is 
\begin{eqnarray}
(D_{t_1}^2-D_{t_2})Z_p(t) \cdot Z_{p-1}(t)\,
=\,
0\,.
\end{eqnarray}
This is the first equation of the modified KP hierarchy.

\subsection{Transfer matrix approach}
The main tool we use in the text is 
the transfer matrix formulation  
of random plane partition.  
The hamiltonian picture is hinted from 
the interlace relations (\ref{interlace relations}), 
which state that plane partitions 
are certain evolutions of partitions 
by the discretized time $m$.  
In particular, 
the transfer matrix formulation \cite{Okounkov-Reshetikhin}
makes it possible to express 
the partition function (\ref{Z_p(t)}) 
in terms of two-dimensional conformal field theory 
(2$d$ free fermions).

Let 
$\psi(z)=\sum_{m \in \mathbb{Z}}\psi_mz^{-m-1}$ 
and 
$\psi^*(z)=\sum_{m \in \mathbb{Z}}\psi^*_mz^{-m}$ 
be complex fermions with the anti-commutation relations, 
$\left\{ \psi_m,\psi^*_n \right\}=\delta_{m+n,0}$ 
and 
$\left\{ \psi_m,\psi_n \right\}=\left\{ \psi_m^*,\psi_n^* \right\}=0$. 
The Noether current of the $U(1)$ rotation is given by   
\begin{eqnarray}
J(z)=\,
: \psi(z)\psi^*(z):\, 
=\,
\sum_{m \in \mathbb{Z}}z^{-m-1}J_m\,, 
\label{J_m}
\end{eqnarray}
where $:~:$ denotes the normal ordering of fermions 
that is defined by 
\begin{eqnarray}
\psi(z)\psi^*(w)=\,
:\psi(z)\psi^*(w):+\frac{1}{z-w}\,,  
\hspace{5mm}
|z| > |w|\,. 
\label{normal ordering}
\end{eqnarray}

It is well known that partitions are realized 
as states of the fermion Fock space. 
In particular, 
charged partitions are realized 
by the states of the same $U(1)$ charges. 
For a charged partition 
$(\lambda,p)$,  
the corresponding state is 
\begin{eqnarray}
|\lambda;p \rangle 
=\,
\prod_{i=1}^{\infty} 
  \psi_{i-\lambda_i-1-p} 
  \psi^*_{-i+1+p}\,
|p\rangle\,, 
\label{charged partition state}
\end{eqnarray}
where 
$|p\rangle$ 
denotes the Dirac sea having the $U(1)$ charge $p$,  
and is defined by the conditions  
\begin{eqnarray}
\psi_m |p\rangle = 0 
&&
\mbox{for~ $\forall\, m \geq -p$}\,, 
\nonumber \\
\psi_m^* |p\rangle = 0 
&&
\mbox{for~ $\forall\, m \geq p+1$}\,. 
\end{eqnarray}

Using the realization (\ref{charged partition state}), 
it can be seen that the function (\ref{Phi_k})  
corresponds to the following fermion bilinear operator. 
\begin{eqnarray}
H_k
&\equiv& 
\sum_{m \in \mathbb{Z}}
q^{km}:\psi_{-m}\psi^*_{m}:\,.  
\label{H_k}
\end{eqnarray}
Actually, the potential function (\ref{Phi_k}) 
is reproduced as 
\begin{eqnarray}
H_k|\lambda;p\rangle 
=
\Phi_k(\lambda,p)|\lambda; p\rangle\,. 
\label{H_k vs Phi_k}
\end{eqnarray}
These operators are commutative. 
The following combination reproduces the potential (\ref{Phi_(t,p)}).  
\begin{eqnarray}
H(t)=\sum_{k=1}^{\infty}t_kH_k\,. 
\label{H(t)}
\end{eqnarray}

The transfer matrices \cite{Okounkov-Reshetikhin} 
are vertex operators of the following forms.
\begin{eqnarray}
\Gamma_{+}(m) 
&=& 
\exp 
  \Bigl(   
      \sum_{k=1}^{+\infty}  
      \frac{1}{k}q^{-k(m+\frac{1}{2})}J_{k}   
  \Bigr)\,, 
\label{Gamma_+(m)} \\
\Gamma_{-}(m) 
&=&
\exp 
  \Bigl(
     \sum_{k=1}^{+\infty}
     \frac{1}{k}q^{k(m+\frac{1}{2})}J_{-k}
  \Bigr)\,, 
\label{Gamma_-(m)}
\end{eqnarray}
where 
$J_{\pm k}$ are the modes of the $U(1)$ current.  
The matrix elements 
between partitions of different charges 
always vanishes, 
while those between partitions of the same charge are 
\begin{eqnarray}
\langle \lambda;p | \Gamma_{+}(m) |\mu;p\rangle 
&=&
\left\{
 \begin{array}{cc}
   q^{-(m+\frac{1}{2})(|\mu|-|\lambda|)} & \mbox{$\lambda\prec\mu$} \\
   0 & \mbox{otherwise}\,.
 \end{array} 
\right. 
\label{Gamma_+ matrix element}\\
\langle \mu;p |\Gamma_{-}(m)|\lambda;p\rangle 
&=&
\left\{
 \begin{array}{cc}
    q^{(m+\frac{1}{2})(|\mu|-|\lambda|)} & \mbox{$\mu\succ\lambda$} \\
    0 & \mbox{otherwise}\,.
 \end{array} 
\right. 
\label{Gamma_- matrix element}
\end{eqnarray}
By comparing these formulas with 
the interlace relations (\ref{interlace relations}), 
we see that $\Gamma_{\pm}(m)$ describes 
the evolutions of partitions. 
More precisely, 
the evolution at a negative time $m \le -1$ 
is given by $\Gamma_+(m)$,  
while the evolution at a nonnegative time $m \ge 0$ 
is by $\Gamma_-(m)$.

Taking the hamiltonian picture of plane partitions, 
the partition function (\ref{Z_p(t)}) can be reproduced 
in the transfer matrix formulation. 
Actually, 
following the same steps 
as we translated the partition function 
to the $q$-deformed random partition (1.10), 
but using the transfer matrices 
in place of the Schur functions, 
the partition function (\ref{Z_p(t)}) 
is eventually expressed as  
\begin{eqnarray}
Z_p(t)&=& 
\langle p|\, 
G_+\, 
e^{H(t)}\, 
G_-\,
|p \rangle\,, 
\label{Z_p(t) CFT}
\end{eqnarray}
where $G_{\pm}$ are the propagators 
those are responsible respectively to 
the negative time evolutions and the nonnegative time evolutions 
of partitions. 
These are the operators given by the following infinite products.
\begin{eqnarray}
G_+ 
&\equiv& 
\prod_{m=-\infty}^{-1}\Gamma_+(m)\,, 
\label{G_+}\\
G_- 
&\equiv& 
\prod_{m=0}^{+\infty}\Gamma_-(m)\,. 
\label{G_-}
\end{eqnarray}

\subsection{Quantum torus Lie algebra and random plane partition}

Starting from the expression (\ref{Z_p(t) CFT}) 
of the partition function (\ref{Z_p(t)}), 
we prove in the text that 
the series of the partition functions (\ref{tau(t;p)=Z_p}) 
is a tau function of the one-dimensional Toda hierarchy. 
Remarkable relations between random plane partition 
and quantum torus Lie algebra are revealed 
in the course of the proof. 
Throughout the text, 
we take the perspective that 
such a quantum Lie algebra is 
a hidden symmetry of random plane partition.

We realize the quantum torus Lie algebra 
in terms of the complex fermions.  
Using this realization, 
we can regard the operators $H_k$ 
as a commutative sub-algebra of the quantum torus Lie algebra. 
The adjoint actions of the propagators $G_{\pm}$ 
on the Lie algebra 
generate automorphisms of the algebra. 
By taking advantage of such automorphisms, 
we provide a proof of the statement. 
Actually, among such automorphisms,  
we pay special attention to the shift symmetry 
that is the automorphism generated by 
the adjoint action of the product $G_-G_+$ 
or equivalently $G_+G_-$. 
By utilizing this symmetry, 
we can eventually express 
the partition function 
in the form (\ref{Z_p(t) Toda}). 
In particular, 
the element of $GL(\infty)$ 
in the formula (\ref{Z_p(t) Toda}) 
is given by 
\begin{eqnarray}
{\bf g}_{\star}
\,\equiv\,
q^{\frac{W}{2}} 
(G_-G_+)^2
q^{\frac{W}{2}}\,, 
\label{g_star}
\end{eqnarray}
where $W$ is a generator of the $W_{\infty}$-algebra 
of the following form: 
\begin{eqnarray}
W
\,\equiv\,  
\sum_{m \in \mathbb{Z}}m^2 :\psi_{-m}\psi^*_{m}:\,. 
\label{W_0(3)}
\end{eqnarray} 
Using this formula, 
we finally confirm the statement by showing that 
${\bf g}_\star$ actually realizes a solution of 
the one-dimensional Toda hierarchy.

Taking the formula (\ref{Z_p(t) Toda}), 
Virasoro/W-constraints on the partition function (\ref{Z_p(t)}) 
and the tau function (\ref{tau(t;p)=Z_p}) can be obtained 
from the transformation of the $W_\infty$-algebra 
by the adjoint action of ${\bf g}_\star$. 
In the same way, 
the transformation of the quantum torus Lie algebra by 
the adjoint action of ${\bf g}_\star$  
gives rise to quantum torus analogues 
of the Virasoro/W-constraints 
on the partition function (\ref{Z_p(t)}) 
and the tau function (\ref{tau(t;p)=Z_p}). 
As we argue subsequently, 
such quantum torus analogues of the Virasoro/W-constraints 
are also obtainable in five-dimensional 
$\mathcal{N}=1$ supersymmetric gauge theories 
and certain topological string amplitudes. 
These are reported in a separate publication \cite{Nakatsu-Takasaki}.

\subsection{Integrability 
of 5$d$ $\mathcal{N}=1$ SUSY gauge theories 
from random plane partition}

Random plane partition has a significant relation with 
five-dimensional $\mathcal{N}=1$ supersymmetric gauge theories 
\cite{MNTT1,MNTT2,MNNT,MN}. 
Our study of the integrable structure of random plane partition 
is motivated by a quest for the integrable structure of 
five-dimensional $\mathcal{N}=1$ supersymmetric gauge theories 
and topological strings.

\subsubsection{Random plane partition and 
5$d$ $\mathcal{N}=1$ SUSY gauge theories}

Nekrasov's functions for five-dimensional $SU(N)$ gauge theories
\cite{Nekrasov, Nekrasov-Okounkov} 
are interpreted as partition functions 
of random plane partition \cite{MNTT1}. 
Actually, 
the original partition function (\ref{Z}) 
reproduces Nekrasov's function for 
five-dimensional $U(1)$ gauge theory, 
by adding a simple chemical potential 
$Q^{|\pi(0)|}$ to the statistical weight. 
In the transfer matrix approach, 
that partition function becomes
\begin{eqnarray}
Z^{5d\,U(1)}&=& 
\langle 0|\,G_+\,Q^{L_0}\,G_-|0 \rangle\,, 
\label{Z_5dU(1)}
\end{eqnarray}
where 
$\displaystyle{L_0 \equiv \sum_{m \in \mathbb{Z}}m:\psi_{-m}\psi^*_m:}$.
The above matrix element is easily computed and leads 
\begin{eqnarray}
Z^{5d\,U(1)}&=&
\prod_{n=1}^{+\infty}\frac{1}{(1-Qq^n)^n}\,. 
\label{Z_5dU(1)2}
\end{eqnarray}
Indeterminates $q,Q$ in the right hand side of this formula 
must be interpreted in terms of the gauge theory parameters 
to reproduce Nekrasov's function for 
the five-dimensional gauge theory. 
The gauge theory lives on $\mathbb{R}^4 \times S^1_{R}$, 
where $R$ denotes the radius of the circle, 
and has the dynamical scale $\Lambda$. 
The indeterminates are identified with 
these parameters by the relations 
\begin{eqnarray}
q=e^{-R \hbar}\,, 
\hspace{8mm}
Q=(R\Lambda)^2\,. 
\label{R,Lambda}
\end{eqnarray}

By shrinking the circle to a point, 
from Nekrasov's functions for five-dimensional gauge theories, 
one obtains the four-dimensional versions. 
Considering the relations (\ref{R,Lambda}), 
this indicates that 
the partition function (\ref{Z_5dU(1)}) 
is a $q$-analogue of the four-dimensional version. 
To see this, 
we note that the four-dimensional limit 
of the right hand side of (\ref{Z_5dU(1)}) 
is obtained by employing  
the dressed propagators, 
$Q^{-\frac{1}{2}L_0}G_+Q^{\frac{1}{2}L_0}$ 
and  
$Q^{\frac{1}{2}L_0}G_-Q^{-\frac{1}{2}L_0}$. 
Actually, 
taking the relations (\ref{R,Lambda}), 
these dressed operators become nonsingular 
at the limit $R \rightarrow 0$, 
and eventually give the following operators. 
\begin{eqnarray}
\lim_{R\,\rightarrow\,0}\,
Q^{-\frac{1}{2}L_0}G_+\,Q^{\frac{1}{2}L_0}\,
&=&
\Lambda^{-L_0}\,e^{\frac{1}{\hbar}J_1}\,\Lambda^{L_0}\,, 
\label{4d limit G_+}
\\
\lim_{R\,\rightarrow\,0}\,
Q^{\frac{1}{2}L_0}\,G_-\,Q^{-\frac{1}{2}L_0}\,
&=& 
\Lambda^{L_0}\,e^{\frac{1}{\hbar}J_{-1}}\,\Lambda^{-L_0}\,. 
\label{4d limit G_-}
\end{eqnarray}
By using these formulas, 
one obtains  
\begin{eqnarray}
\lim_{R\,\rightarrow\,0}
\langle 0|\,G_+\,Q^{L_0}\,G_-|0 \rangle\, 
&=&
\lim_{R\,\rightarrow\,0}
\langle 0|                        
Q^{-\frac{1}{2}L_0}G_+Q^{\frac{1}{2}L_0}\, 
Q^{\frac{1}{2}L_0}G_-Q^{-\frac{1}{2}L_0}
|0 \rangle\, 
\nonumber \\
&=& 
\langle 0|\, 
e^{\frac{1}{\hbar}J_1}\,
\Lambda^{2L_0}\, 
e^{\frac{1}{\hbar}J_{-1}}\, 
|0\rangle\,.  
\label{Z_4dU(1)}
\end{eqnarray}
The right hand side of this formula is nothing but 
Nekrasov's function for four-dimensional $U(1)$ gauge theory.

\subsubsection{Integrable structure 
of 5$d$ $\mathcal{N}=1$ SUSY gauge theories}

The operator $H_k$ has a counterpart in five-dimensional gauge theories.  
It corresponds to the Wilson loop operator 
encircling the circle $k$ times \cite{5d observable}.
\begin{eqnarray}
O_k=
\mbox{Tr}\, 
\left\{
P e^{\oint dt\, A_4+i\varphi}
\right\}^k\,, 
\label{O_k}
\end{eqnarray}
where $A_4$ and $\varphi$ denote respectively 
the fifth component of the gauge field 
and the real scalar field 
in the vector multiplet.  
Generating function of the correlators among  
these observables becomes thereby 
the following analogue of (\ref{Z_p(t) CFT}).  
\begin{eqnarray}
Z^{5d\,U(1)}_p(t)
\,=\,
\langle p|\, 
G_+\, 
e^{H(t)}\, 
Q^{L_0}
G_-\,
|p \rangle\,. 
\label{Z_p(t)5dU(1)}
\end{eqnarray}

In the same way as we stated on (\ref{tau(t;p)=Z_p}), 
the series of the generating functions 
(\ref{Z_p(t)5dU(1)}) also 
gives rise to a tau function of the one-dimensional Toda hierarchy. 
In particular, 
the tau function has the following expression 
analogous to (\ref{Z_p(t) Toda}).
\begin{eqnarray}
\tau^{5d\,U(1)}(t;p)\,=\,
e^{\sum_{k=1}^{\infty}\frac{t_kq^k}{1-q^k}}\,\,
\langle p |\,
e^{\frac{1}{2}\sum_{k=1}^{\infty}(-)^kt_kJ_k}
\,\,
{\bf g}_{\star}^{5d\,U(1)}
\,\, 
e^{\frac{1}{2}\sum_{k=1}^{\infty}(-)^kt_kJ_{-k}}\,\, 
|p \rangle\,,
\label{tau_p(t) 5dU(1)}
\end{eqnarray}
where ${\bf g}_{\star}^{5d\,U(1)}$ is the element 
of $GL(\infty)$ given by 
\begin{eqnarray}
{\bf g}_{\star}^{5d\,U(1)}
\,\equiv\,
q^{\frac{W}{2}} 
(G_-G_+)\,
Q^{L_0}\,
(G_-G_+)\,
q^{\frac{W}{2}}\,.
\label{g_star 5dU(1)}
\end{eqnarray}
We note that 
this formula is easily generalized 
to the $SU(N)$ gauge theory, 
where the series of the corresponding generating functions 
of the correlators becomes 
a tau function of the one-dimensional Toda hierarchy.

In four-dimensions,  
such an integrable structure of 
$\mathcal{N}=2$ supersymmetric gauge theories 
has been found out 
\cite{Nekrasov-Marshakov,Nekrasov-Losev} 
among the generating functions of 
the higher Casimir operators 
$\mbox{Tr}\phi^k$, where $\phi$ is the complex scalar 
in the vector multiplet, 
corresponding to $A_4+i\varphi$ 
in five-dimensional theories.  
One might expect that 
the tau function (\ref{tau_p(t) 5dU(1)}) 
is a $q$-analogue of the tau function 
of the four-dimensional theory, 
just like 
the partition function (\ref{Z_5dU(1)})
is the $q$-analogue.  
However, this is not the case. 
Relation between these two integrable structures 
is not straightforward. 
The subtlety can be found, for instance, 
in that the tau function (\ref{tau_p(t) 5dU(1)}) 
becomes trivial at the four-dimensional limit, 
owing to the degeneration of all the observables (\ref{O_k}).  
Actually, 
when $R$ is nearly zero, 
the generating function (\ref{Z_p(t)5dU(1)}) behaves as 
\begin{eqnarray}
Z^{5d\,U(1)}_p(t)\,
\sim\, 
R^{p(p+1)}e^{p\sum_{k=1}^{\infty}t_k}\,
\langle 0|\, 
e^{\frac{1}{\hbar}J_1}\,
\Lambda^{2L_0}\, 
e^{\frac{1}{\hbar}J_{-1}}\, 
|0\rangle\,.
\label{Z_p(t)5dU(1) 4d limit}
\end{eqnarray}

\subsection{Integrability of topological string amplitudes 
from random plane partition}

\begin{figure}[ht]
\begin{center}
\includegraphics[scale=.6]{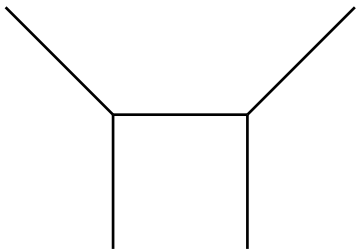}
\end{center}
\caption{The diagram for 
$\mathcal{O}\oplus \mathcal{O}(-2) \rightarrow  \mathbb{C}\mbox{P}^1$.}
\label{diagram for U(1) geometry}
\end{figure}

In addition to the gauge theory interpretation, 
the partition function (\ref{Z_5dU(1)}) 
has an interpretation as all genus $A$-model topological 
string amplitude on 
$\mathcal{O}\oplus \mathcal{O}(-2) \rightarrow  \mathbb{C}\mbox{P}^1$. 
It is a non-compact toric Calabi-Yau threefold, 
often called a local geometry. 
The toric description is given by a fan $\Delta$ 
consisting of rational cones of dimensions $\leq 3$ 
on $\mathbb{R}^3$. 
The topological vertex 
\cite{Iqbal, Aganagic-Klemm-Marino-Vafa} 
is a diagrammatical method to compute all genus $A$-model topological 
string amplitudes for such local geometries. 
The diagram can be drawn from a polyhedron 
which is obtained by taking duals  
to the cones in $\mathbb{R}^3$.
For the local geometry 
$\mathcal{O}\oplus \mathcal{O}(-2) \rightarrow  \mathbb{C}\mbox{P}^1$, 
the relevant diagram is depicted in Figure \ref{diagram for U(1) geometry}. 
The topological vertex computation based on the diagram gives 
all genus $A$-model topological string amplitude on 
$\mathcal{O}\oplus \mathcal{O}(-2) \rightarrow  \mathbb{C}\mbox{P}^1$ 
as 
\begin{eqnarray}
\mathcal{A}^{\mathcal{O}\oplus \mathcal{O}(-2) }_{string}
=\prod_{n=1}^{+\infty}\frac{1}{(1-e^{-a}e^{-ng_{st}})^{n}}\,, 
\label{amplitude U(1) geometry}
\end{eqnarray}
where $a$ denotes 
the K$\ddot{\mbox{a}}$hler volume of the base $\mathbb{C}\mbox{P}^1$, 
and $g_{st}$ is the string coupling constant. 
Comparing two formulas (\ref{Z_5dU(1)2}) and (\ref{amplitude U(1) geometry}), 
one sees that 
the partition function (\ref{Z_5dU(1)}) becomes 
all genus $A$-model topological string amplitude on 
$\mathcal{O}\oplus \mathcal{O}(-2) \rightarrow  \mathbb{C}\mbox{P}^1$, 
by the following identification of the parameters. 
\begin{eqnarray}
q=e^{-g_{st}}\,, 
\hspace{8mm}
Q=e^{-a}\,. 
\label{a,g_st}
\end{eqnarray}
%
%

As is the case of the five-dimensional gauge theory, 
we generalize the amplitude 
including the operators $H_k$ in it,  
anticipating that 
a wound Euclidean brane along the $M$-theory circle
\footnote{We thank Y. Hyakutake 
for suggesting such a possibility to us.} 
corresponds to the observable $O_k$ of the gauge theories. 
We physically conjecture such a generalization by 
\begin{eqnarray}
\mathcal{A}^{\mathcal{O}\oplus \mathcal{O}(-2) }_{string}(t;p)
\,=\,
\langle p|\, 
G_+\, 
e^{H(t)}\, 
Q^{L_0}
G_-\,
|p \rangle\,, 
\label{A(t;p)_U(1)geometry}
\end{eqnarray}
where the right hand side of this equation is same as 
the formula (\ref{Z_p(t)5dU(1)}) 
but with the different interpretation (\ref{a,g_st}). 
The series of the generating functions 
(\ref{A(t;p)_U(1)geometry}) gives 
the same tau function as is the case of 
the five-dimensional $U(1)$ gauge theory.

It seems rather rare that the one-dimensional Toda hierarchy 
shows up as an integrable structure of topological strings. 
One of such rare cases is the topological sigma model 
(in other words, the Gromov-Witten invariants) 
of $\mathbb{C}P^1$ 
\cite{Eguchi-Hori-Yang, Eguchi-Hori-Xiong, Getzler, Pandharipande, 
OK1, Givental}. 
On the other hand, the ``relative'' or ``equivariant'' 
versions of those Gromov-Witten invariants 
have a different integrable structure, 
namely, the two-dimensional Toda hierarchy \cite{OK1,OK2}. 
It is remarkable that substantially the same quantum torus 
Lie algebra is used in the work of Okounkov and Pandharipande 
\cite{OK2}, but we have been unable to see whether there 
is a deep connection with our work.  To obtain the generators 
$V^{(k)}_m$ of our quantum torus Lie algebra, one has to 
specialize the parameter $z$ of Okounkov and Pandharipande's 
operators ${\mathcal E}_m(z)$ to $e^z = q^k$;  such powers of $q$ 
apparently play no role in the work of Okounkov and Pandharipande.

The two-dimensional Toda hierarchy 
is also known to arise in the generating function 
$\tau(x,\bar{x})$
$=\sum_{\lambda,\mu}s_\lambda(x)s_\mu(\bar{x})c_{\lambda\mu\bullet}$ 
of the two-legged topological vertex $c_{\lambda\mu\bullet}$ 
\cite{Zhou}. 
By changing variables from 
$x,\bar{x}$ to $T_k 
= \frac{1}{k}\sum_i x_i^k$, $\bar{T}_k 
= - \frac{1}{k}\sum_i \bar{x}_i^k$, 
$\tau(x,\bar{x})$ coincides 
with the special part $\tau^{2\,Toda}(T,\bar{T},0)$ 
of a tau function $\tau^{2\,Toda}(T,\bar{T},p)$ 
of the two-dimensional Toda hierarchy.  
As Zhou pointed out, 
the tau function $\tau^{2\,Toda}(T,\bar{T},p)$ 
has a fermionic representation 
in terms of an element $g$ of $GL(\infty)$ 
of the form 
\begin{eqnarray}
  g = q^{\frac{K}{2}}G_{+}G_{-}q^{\frac{K}{2}},
\end{eqnarray}
where $K = \sum_m (m-\frac{1}{2})^2 :\psi_{-m}\psi^*_m:$. 
Thus the building blocks of this tau function 
are similar to the foregoing partition functions 
of $U(1)$ gauge theory. 
The difference between $K$ and $W$ is almost negligible, 
and can be absorbed by rescaling of $T_k,\bar{T}_k$ 
and an exponential prefactor.

\subsubsection{Emergence of geometry from condensation}
It is amazing to comment on a possibility of 
emergence of another geometry from the condensation 
in a local geometry.

Consider the generating function (\ref{A(t;p)_U(1)geometry}) for 
the case of $p=0$, and write it simply as  
$\mathcal{A}^{\mathcal{O}\oplus \mathcal{O}(-2) }_{string}(t)$
$\equiv$ 
$\mathcal{A}^{\mathcal{O}\oplus \mathcal{O}(-2) }_{string}(t;p=0)$. 
As explained in the beginning of Section 4, 
(\ref{A(t;p)_U(1)geometry}) has another representation 
of the following form. 
\begin{eqnarray}
\mathcal{A}^{\mathcal{O}\oplus \mathcal{O}(-2) }_{string}(t)
\,=\,
e^{\sum_{k=1}^{\infty}\frac{t_kq^k}{1-q^k}}\,\,
\langle 0 |\,
e^{\sum_{k=1}^{\infty}(-)^kt_kJ_k}
\,\,
{\bf g}_{\star}^{5d\,U(1)}
\, 
|0 \rangle\,.
\label{A(t;0)_U(1)geometry}
\end{eqnarray}
This becomes a tau function of the KP hierarchy.

The generating function has 
a factorization in terms of the skew Schur functions. 
Matrix element in the right hand side of 
formula (\ref{A(t;0)_U(1)geometry}) 
is factored by plugging the unity 
$1=\sum_{p \in \mathbb{Z}}\sum_{\lambda}|\lambda; p\rangle \langle \lambda;p|$  so that it divides ${\bf g}^{5d\,U(1)}_\star$ into two parts.  
Owing to the charge conservation, 
the sum over $p$ truncates to $p=0$. 
Writing $|\lambda;0\rangle$ simply as $|\lambda \rangle$,  
we can obtain 
\begin{eqnarray}
\langle 0|\,
e^{\sum_{k=1}^{\infty}(-)^kt_kJ_k}\,\,
{\bf g}_{\star}^{5d\,U(1)}\,
|0 \rangle
\,=\,
\sum_{\lambda}\,
\langle 0|\,
e^{\sum_{k=1}^{\infty}(-)^kt_kJ_k}\,\,
q^{\frac{W}{2}}G_-\,
|\lambda \rangle 
\times
\langle \lambda |\,
G_+Q^{L_0}G_-\,
|0 \rangle\,.
\label{factorization of A 1}
\end{eqnarray}
Matrix elements in the right hand side of this equation 
are expressed in terms of the skew Schur functions 
as follows.  
\begin{eqnarray}
\langle 0|\,
e^{\sum_{k=1}^{\infty}(-)^kt_kJ_k}\,\,
q^{\frac{W}{2}}G_-\,
|\lambda \rangle 
&=&
\sum_{\mu}\,
q^{\frac{\kappa(\mu)}{2}+\frac{|\mu|}{2}}\,
s_{\mu}(x)s_{\mu/\lambda}(q^{\rho})\,, 
\label{1st factor}
\\
\langle \lambda |\,
G_+Q^{L_0}G_-\,
|0 \rangle 
&=&
\frac{Q^{|\lambda|}s_{\lambda}(q^{\rho})}
{\prod_{n=1}^{+\infty}(1-Qq^n)^{n}}\,, 
\label{2nd factor}
\end{eqnarray}
where variables are changed from $t$ to $x$ 
by $t_k=\frac{1}{k}\sum_{i}(-x_i)^k$. 
Combining these two formulas, 
the right hand side of formula (\ref{A(t;0)_U(1)geometry}) 
is expressed in terms of the skew Schur functions. 
Therefore, 
the factorization of the generating function, 
normalized by the original $A$-model topological string amplitude, 
can be written as  
\begin{eqnarray}
\frac{\mathcal{A}^{\mathcal{O}\oplus \mathcal{O}(-2) }_{string}(t)}
{\mathcal{A}^{\mathcal{O}\oplus \mathcal{O}(-2) }_{string}(0)} 
\,=\, 
e^{\sum_{k=1}^{\infty}\frac{t_kq^k}{1-q^k}}\,\,
\sum_{\lambda}\,
Q^{|\lambda|}s_{\lambda}(q^{\rho})
\left\{
\sum_{\mu}\,
q^{\frac{\kappa(\mu)}{2}+\frac{|\mu|}{2}}\,
s_{\mu}(x)s_{\mu/\lambda}(q^{\rho})
\right\}\,.
\label{factorization A 2}
\end{eqnarray}

We examine the condensation by choosing the coupling constants $t$ 
at certain values. In particular, we take the following values.
\begin{eqnarray}
t_{k}^\star 
&=&
\frac{q^{\frac{k}{2}}}{k(1-q^k)}\,, 
\hspace{8mm}k=1,2,...\,.
\label{t_k_star}
\end{eqnarray} 
Owing to the identification (\ref{a,g_st}), 
$t^\star$ behave $\sim g_{st}^{-1}$ when 
$g_{st}$ is nearly zero and therefore 
a possible condensation becomes 
nonperturbative in $IIA$ superstrings.  
However, one can compute the right hand side of 
formula (\ref{factorization A 2}). 
The computation yields eventually 
the amplitude as 
\begin{eqnarray}
\frac{\mathcal{A}^{\mathcal{O}\oplus \mathcal{O}(-2) }_{string}(t^\star)}
{\mathcal{A}^{\mathcal{O}\oplus \mathcal{O}(-2) }_{string}(0)} 
&=& 
\prod_{n=1}^{+\infty}
(1-Qq^{n+\frac{1}{2}})^n\,. 
\label{A(t_star)}
\end{eqnarray}
The right hand side of this formula coincides 
with all genus $A$-model topological string amplitude 
on the resolved conifold 
$\mathcal{O}(-1)\oplus \mathcal{O}(-1) \rightarrow  \mathbb{C}\mbox{P}^1$, 
where the K$\ddot{\mbox{a}}$hler volume of the base $\mathbb{C}\mbox{P}^1$ is 
$a+\frac{1}{2}g_{st}$.

\begin{figure}[ht]
\begin{center}
\includegraphics[scale=.6]{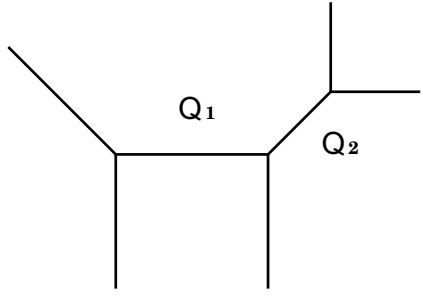}
\end{center}
\caption{The diagram for two conifolds.}
\label{diagram for two conifolds}
\end{figure}

Emergence of the resolved conifold 
in (\ref{A(t_star)}) seems mysterious. 
However, this can be explained as follows. 
Let us consider the local geometry of coupled conifolds. 
Coupled conifolds can be obtained by patching torically 
together the foregoing local geometry and $\mathbb{C}^2$. 
The diagram of coupled conifolds is depicted 
in Figure \ref{diagram for two conifolds}, 
where $Q_{1,2}$ are the K$\ddot{\mbox{a}}$hler parameters 
attached to the internal edges of the diagram.
Each internal edges corresponds to $\mathbb{C}P^1$. 
The K$\ddot{\mbox{a}}$hler parameters are given 
by $Q_{1,2}=e^{-a_{1,2}}$, where $a_{1,2}$ 
denote the K$\ddot{\mbox{a}}$hler volumes of 
the corresponding $\mathbb{C}P^1$s.
Based on this diagram, 
the topological vertex computation  
yields the all genus $A$-model topological string amplitude as 
\begin{eqnarray}
\mathcal{A}^{two\,\, conifolds}_{string}
&=&
\frac{\prod_{n=1}^{+\infty}(1-Q_1Q_2q^{n})^n 
\cdot \prod_{n=1}^{+\infty}(1-Q_2q^{n})^n }
{\prod_{n=1}^{+\infty}(1-Q_1q^{n})^n }\,. 
\label{amplitude on two conifolds}
\end{eqnarray}
%
%

It is remarkable that 
the topological string amplitude (\ref{amplitude on two conifolds}) 
appears,
by tuning the K$\ddot{\mbox{a}}$hler volumes $a_{1,2}$, 
as a building block of the generating function 
(\ref{A(t;0)_U(1)geometry}) at $t=t^\star$.  
Actually, 
the matrix element  
$\langle 0|\,
e^{\sum_{k=1}^{\infty}(-)^kt_kJ_k}\,\,
{\bf g}_{\star}^{5d\,U(1)}\,|0 \rangle$ 
is evaluated by using formulas 
(\ref{1st factor}),(\ref{2nd factor}) 
and becomes eventually as follows.
\begin{eqnarray}
\left.
\langle 0|\,
e^{\sum_{k=1}^{\infty}(-)^kt_kJ_k}\,\,
{\bf g}_{\star}^{5d\,U(1)}\,|0 \rangle 
\right|_{t=t^\star}
&=&
\mathcal{A}^{two\,\, conifolds}_{string}\,, 
\label{matrix element=A_two conifolds}
\end{eqnarray}
where the K$\ddot{\mbox{a}}$hler volumes $a_{1,2}$ 
in the right hand side of this formula are 
\begin{eqnarray}
a_1=a\,, 
\hspace{8mm}
a_2=\frac{1}{2}g_{st}\,. 
\label{a_1,a_2}
\end{eqnarray}
This formula indicates that the condensation (\ref{t_k_star}) 
changes the original geometry into 
two-conifolds with two-cycles 
having the K$\ddot{\mbox{a}}$hler volumes (\ref{a_1,a_2}),  
and that the ratio (\ref{A(t_star)}) counts 
the worldsheet instantons 
wrapping the two-cycles simultaneously. 
The further issue will be reported in \cite{Nakatsu-Takasaki}.

\subsubsection*{\underline{Organization of the article}}
The purpose of this article is to show 
that the series of the partition functions (\ref{tau(t;p)=Z_p}) 
is a tau function of one-dimensional Toda hierarchy. 
We start Section 2 
with giving the realization of quantum torus Lie algebra 
by using the complex fermions.  
In Section 3, 
we argue the automorphisms of the Lie algebra 
generated by the adjoint actions of $G_{\pm}$. 
By using such automorphisms  
we confirm that 
the series of the partition functions (\ref{tau(t;p)=Z_p}) 
satisfy the Toda equation 
(\ref{Toda equation in Hirota form}). 
In Section 4, 
we provide a proof of the statement.

\subsubsection*{\underline{Acknowledgements}}
We are very grateful to T. Tamakoshi 
for his participating the research at the early stage of this work. 
T.N. benefited from discussion with K. Tsuda and Y. Noma. 
K.T. is also grateful to M. Mulase 
for fruitful discussion on a related issue. 
Finally we thank the referees for 
useful comments and helpful suggestion. 
K.T is supported in part by Grant-in-Aid for Scientific Research 
No. 18340061 and No. 19540179.


\section{Quantum torus Lie algebra} 

Let $V^{(k)}_m$, 
where $k=0,1,2,\cdots$ and $m \in \mathbb{Z}$, 
be a set of operators which are defined 
by the following generating function 
for each $k$. 
\begin{eqnarray}
:\psi(q^{\frac{k}{2}}z)\psi^*(q^{-\frac{k}{2}}z):
&=&
\sum_{m \in \mathbb{Z}}
z^{-m-1}q^{-\frac{k}{2}}
V^{(k)}_m\,. 
\label{V_m(k)}
\end{eqnarray}
This formula yields 
the following expression of $V^{(k)}_m$.   
\begin{eqnarray}
V^{(k)}_m\, 
=\,
q^{\frac{k}{2}} 
\oint 
\frac{dz}{2\pi i}
z^m 
:\psi(q^{\frac{k}{2}}z)\psi^*(q^{-\frac{k}{2}}z):\,,  
\label{V_m(k) integral}
\end{eqnarray}
where the integral means taking the residue at $z=0$. 
This integral can be evaluated by 
plugging the mode expansion of $\psi(z),\psi^*(z)$ 
into the generating function. 
Thereby, 
the right hand side of equation 
(\ref{V_m(k) integral}) 
is read as 
\begin{eqnarray}
V_m^{(k)}\,
=\,
q^{-\frac{km}{2}}
\sum_{n \in \mathbb{Z}} 
q^{kn}
:\psi_{m-n}\psi^*_n:\,. 
\label{V_m(k) mode}
\end{eqnarray}
The operator $H_k$ (\ref{H_k}) 
and the $U(1)$ current $J_m$ (\ref{J_m}) 
are represented as 
\begin{eqnarray}
H_k\,=\,V^{(k)}_0\,,
\hspace{8mm}
J_m\,=\,V^{(0)}_m\,.
\end{eqnarray}
Henceforth, 
taking the viewpoint of integrable systems, 
we call $H_k$ ($k \ge 1$) hamiltonians.

We note that 
the normal ordering in the formula (\ref{V_m(k)})
is redundant when $k \neq 0$. 
Actually, 
the fermion bilinear form 
$-\psi^*(q^{-\frac{k}{2}}z)\psi(q^{\frac{k}{2}}z)$
is regularized solely 
by the point splitting\footnote{We take $0<q<1$.} 
$z \rightarrow  q^{\pm \frac{k}{2}}z$, 
without any normal ordering. 
Effect of the normal ordering is known  
from (\ref{normal ordering}) 
and becomes the subtraction of a finite term as  
\begin{eqnarray}
-\psi^*(q^{-\frac{k}{2}}z) \psi(q^{\frac{k}{2}}z) 
=
:\psi(q^{\frac{k}{2}}z)\psi^*(q^{-\frac{k}{2}}z): 
-\frac{q^{\frac{k}{2}}}{z(1-q^k)}\,. 
\label{normal ordered form}
\end{eqnarray}
Thereby, 
the normal ordering only makes 
the finite gap between $V_0^{(k)}$ and 
that obtained without the normal ordering 
as follows. 
\begin{eqnarray}
V_0^{(k)}
&=& 
-\sum_m q^{km} \psi^*_m\psi_{-m}
+\frac{q^k}{1-q^k}\,. 
\label{V_0(k)}
\end{eqnarray}

The operators $V^{(k)}_m$ satisfy 
quantum torus Lie algebra 
(the sine-algebra \cite{Fairlie}).  
The following commutation relations can be found out. 
\begin{eqnarray}
\Bigl[ V^{(k)}_{m}\,, V^{(l)}_n \Bigr] 
= 
(q^{\frac{lm-kn}{2}}-q^{-\frac{lm-kn}{2}}) 
\Bigl( 
V^{(k+l)}_{m+n}-
\delta_{m+n,0}
\frac{q^{k+l}}{1-q^{k+l}}
\Bigr)\,,  
\label{V-algebra} 
\end{eqnarray}
where 
$k,l=0,1,\cdots$ 
and 
$m,n \in \mathbb{Z}$. 
The commutation relations (\ref{V-algebra}) 
become the standard ones by shifting 
$V^{(k)}_0 \rightarrow V^{(k)}_0-\frac{q^k}{1-q^k}$ 
for $k \neq 0$. 
The hamiltonians $H_k (k \geq 0)$, 
where $H_0\equiv V^{(0)}_0$ is included, 
generate a commutative sub-algebra of 
this Lie algebra. 
The non-negative modes $\bigl\{ J_m \bigr\}_{m \geq 0}$
and the non-positive modes $\bigl\{J_{-m} \bigr\}_{m \geq 0}$
of the $U(1)$ current do as well. 
In addition to these sub-algebras, 
there are two more commutative sub-algebras 
that are generated respectively by 
$\bigl\{ V^{(k)}_k \bigr\}_{k \geq 0}$ 
and 
$\bigl\{ V^{(k)}_{-k} \bigr\}_{k \geq 0}$. 
All these sub-algebras are found to 
relate with one another.

The appearance of quantum torus Lie algebra 
might be unexpected. 
However it can be explained from 
the viewpoint of the sine-algebra \cite{Fairlie}.  
The sine-algebra is obtained from $sl(N)$ 
by taking the large $N$ limit of its trigonometrical basis. 
Let $X,Y$ be two $N \times N$ unitary matrices given by
\begin{eqnarray}  
X=\sum_{i=1}^{N-1}E_{i,i+1}+E_{N,1}\,,
\hspace{4mm} 
Y=\sum_{i=1}^{N}\omega^{i-1}E_{i,i}\,,
\label{XY} 
\end{eqnarray}
where $\omega$ is a $N$-th root of the unity. 
These two matrices satisfy the relation 
$XY=\omega YX$.  
Their non-commutative monomials $X^mY^k$ for 
$0 \le m < N$ and $0 \le k < N$ 
give the trigonometrical 
basis of $sl(N)$. 
The sine-algebra is the Lie algebra of $X^mY^k$ at $N=\infty$  
and is identified with quantum torus Lie algebra. 
It is the Lie algebra 
derived from quantum two-torus (non-commutative two-torus), 
that is, 
an unital algebra with two generators $U,V$ 
satisfying the relation $UV=qVU$. 
Here $q$ is regarded as the non-commutative parameter. 
The trigonometrical basis $X^mY^k$ are transmuted 
to the non-commutative monomials $U^mV^k$. 
Let us normalize them as follows.      
\begin{eqnarray}
v^{(k)}_m=q^{-\frac{km}{2}}U^mV^k\,. 
\label{v_m(k)}
\end{eqnarray}
These normalized ones satisfy 
the commutation relations (\ref{V-algebra}), 
apart from the shift of the zero modes.

The sine-algebra is the algebra of trigonometric basis 
of $sl(\infty)$.  Making use of the embedding $sl(\infty) 
\subset gl(\infty)$, as utilized in (\ref{XY}) for finite $N$, 
a trigonometric basis can be realized in terms of 
$q$-difference operators with respect to $z$.   
Among them, the fundamental operators are $z$ and 
$q^{-zd/dz} = \exp(-\log q z\frac{d}{dz})$, 
which are the counterparts of $X$ and $Y$.  
These two operators satisfy the relation 
$z q^{-zd/dz} = q q^{-zd/dz}z$. 
The normalized basis (\ref{v_m(k)}) of the sine-algebra 
correspond to 
\begin{eqnarray}
v^{(k)}_m=q^{-\frac{km}{2}}z^mq^{-kz\frac{d}{dz}}\,. 
\label{v_m(k)z}
\end{eqnarray} 
The operators $V_m^{(k)}$ are nothing but the second-quantizations 
of $v^{(k)}_m$ by means of the fermions.  
\begin{eqnarray}
V^{(k)}_m=
\oint 
\frac{dz}{2\pi i}
:\psi(z)v^{(k)}_m \psi^*(z):\,. 
\label{def of V 4}
\end{eqnarray}

Since we have restricted $k \ge 0$, 
the operators $V_m^{(k)}$ or 
the basis (\ref{v_m(k)z}) 
generate only the half of 
quantum torus Lie algebra. 
Needless to say, 
the operators for $k<0$ are obtainable from 
the generating function (\ref{V_m(k)}) by 
choosing $k$ as such, 
however those play no role in this article. 
The above half is, so to speak, 
a quantum cylinder Lie algebra 
to be obtained from a quantum cylinder. 
In the context of random plane partition, 
this quantum cylinder becomes   
a classical cylinder $\mathbb{C}^*$ 
at the thermodynamic limit or the semi-classical limit; 
the Seiberg-Witten hyper-elliptic curves of 
five-dimensional $\mathcal{N}=1$  supersymmetric 
gauge theories emerge as the double cover 
of the cylinder \cite{MNTT2, MN}. 
See \cite{MNTT2, MN} for the details.


\section{One-dimensional Toda chain}
In this section we show that 
the series of the partition functions (\ref{tau(t;p)=Z_p}) 
provides a solution of the one-dimensional Toda chain. 
We will identify the partition functions with  
dynamical variables $\phi_p$ on a $\mathbb{Z}$-lattice by 
\begin{eqnarray}
e^{\phi_p}\equiv 
\frac{\tau(t;p+1)}{\tau(t;p)}
=
q^{(p+1)^2} 
\frac{Z_{p+1}(t)}{Z_p(t)}\,,~~~~~~~
p \in \mathbb{Z}\,.
\label{phi_p}
\end{eqnarray} 
Then, writing $x=t_1$,  
the variables $\phi_p$ satisfy the following Toda equation. 
\begin{eqnarray}
\partial_x^2 \phi_p=e^{\phi_{p+1}-\phi_{p}}-e^{\phi_{p}-\phi_{p-1}}\,,  
~~~~~~~p \in \mathbb{Z}\,. 
\label{Toda equation}
\end{eqnarray}

\subsection{Transformations by adjoint action} 
We argue 
transformations of the generators $V^{(k)}_m$ 
by the adjoint action of 
the propagators $G_{\pm}$. 
For this purpose, 
we conveniently start with the transformations 
of the fermions 
$\psi(z), \psi^*(z)$.
These follow from their rotations 
by the transfer matrices $\Gamma_{\pm}(m)$, 
since $G_{\pm}$ are just the products of $\Gamma_{\pm}(m)$.   
The rotations are easily computed 
by recalling that each modes of the $U(1)$ current 
rotates the fermions as follows.   
\begin{eqnarray}
e^{\theta J_m}\psi(z)e^{-\theta J_m}=e^{\theta z^m}\psi(z)\,, 
~~~~~
e^{\theta J_m}\psi^*(z)e^{-\theta J_m}=e^{-\theta z^m}\psi^*(z)\,, 
~~~~~\theta \in \mathbb{C}\,. 
\label{U(1) on fermions}
\end{eqnarray} 
For instance, the rotation of $\psi(z)$ by $\Gamma_+(m)$ is 
computed as follows. 
\begin{eqnarray}
\Gamma_+(m)\, \psi(z)\,\Gamma_+(m)^{-1}
&=&
\prod_{k=1}^{+\infty}
e^{\frac{1}{k}q^{-k(m+\frac{1}{2})}J_k}\,
\psi(z)\,
\prod_{k=1}^{+\infty}
e^{-\frac{1}{k}q^{-k(m+\frac{1}{2})}J_k}
\nonumber \\
&=& 
e^{\sum_{k=1}^{+\infty}\frac{1}{k}z^kq^{-k(m+\frac{1}{2})}}
\psi(z) 
\nonumber \\
&=&
(1-zq^{-(m+\frac{1}{2})})^{-1}\,
\psi(z)\,. 
\label{example 1}
\end{eqnarray}
By using this formula reiteratively according as (\ref{G_+}),  
we can compute the adjoint action of $G_+$. 
In this way, 
we eventually obtain 
the following transformations of the fermions.  
\begin{eqnarray}
&&
\left\{
\begin{array}{l}
\displaystyle{
G_+\psi(z)(G_+)^{-1}=
\prod_{m=1}^{+\infty}
(1-zq^{m-\frac{1}{2}})^{-1}\,
\psi(z)\,,} 
\\ 
\displaystyle{
G_+\psi^*(z)(G_+)^{-1}=
\prod_{m=1}^{+\infty}
(1-zq^{m-\frac{1}{2}})\,\, 
\psi^*(z)\,.}  
\end{array}
\right. 
\label{G+ on fermions}
\\
&&
\left\{
\begin{array}{l}
\displaystyle{
(G_-)^{-1}\psi(z)G_-=
\prod_{m=0}^{+\infty}
(1-z^{-1}q^{m+\frac{1}{2}})\,\,
\psi(z)\,,}
\\
\displaystyle{
(G_-)^{-1}\psi^*(z)G_-=
\prod_{m=0}^{+\infty}
(1-z^{-1}q^{m+\frac{1}{2}})^{-1}\, 
\psi^*(z)\,.} 
\end{array}
\right. 
\label{G- on fermions} 
\end{eqnarray}

Nextly we examine the transformations of 
the generating function 
(\ref{V_m(k)}) 
of the quantum torus Lie algebra. 
These transformations are obtained 
by using formulas (\ref{G+ on fermions}), (\ref{G- on fermions}) 
taking (\ref{normal ordered form}) into account.
We illustrate the computation for the case of $G_+$. 
\begin{eqnarray}
&&
G_+ 
:\psi(q^{\frac{k}{2}}z)\psi^*(q^{-\frac{k}{2}}z): 
(G_+)^{-1}
\nonumber \\
&&~~~
= 
G_+
\left\{
-\psi^*(q^{-\frac{k}{2}}z)\psi(q^{\frac{k}{2}}z)
+\frac{q^{\frac{k}{2}}}{z(1-q^k)} 
\right\}
(G_+)^{-1}
\nonumber \\
&&~~~
= 
-G_+ 
\psi^*(q^{-\frac{k}{2}}z)(G_+)^{-1}\,  
G_+\psi(q^{\frac{k}{2}}z)(G_+)^{-1}
+
\frac{q^{\frac{k}{2}}}{z(1-q^k)} 
\nonumber \\
&&~~~
= 
-\prod_{m=1}^k(1-zq^{\frac{k+1}{2}-m})\,\, 
\psi^*(q^{-\frac{k}{2}}z)\psi(q^{\frac{k}{2}}z)
+
\frac{q^{\frac{k}{2}}}{z(1-q^k)}\,.  
\label{example 2}
\end{eqnarray}
The last line in (\ref{example 2}) 
can be rewritten 
in terms of the generating function itself. 
We thus obtain 
\begin{eqnarray}
&&
G_+
\left\{ 
:\psi(q^{\frac{k}{2}}z)\psi^*(q^{-\frac{k}{2}}z): 
-\frac{q^{\frac{k}{2}}}{z(1-q^k)}
\right\}
(G_+)^{-1}
\nonumber \\
&&
\hspace{15mm}
=
\prod_{m=1}^k(1-zq^{\frac{k+1}{2}-m})
\left\{ 
:\psi(q^{\frac{k}{2}}z)\psi^*(q^{-\frac{k}{2}}z): 
-\frac{q^{\frac{k}{2}}}{z(1-q^k)}
\right\}\,. 
\label{G+ on psipsi*}
\end{eqnarray}
The similar computation goes as well for the case of $G_-$ and gives  
\begin{eqnarray}
&&
(G_-)^{-1}
\left\{ 
:\psi(q^{\frac{k}{2}}z)\psi^*(q^{-\frac{k}{2}}z): 
-\frac{q^{\frac{k}{2}}}{z(1-q^k)}
\right\}
G_-
\nonumber \\
&&
\hspace{15mm}
=
\prod_{m=1}^k(1-z^{-1}q^{-\frac{k+1}{2}+m})
\left\{ 
:\psi(q^{\frac{k}{2}}z)\psi^*(q^{-\frac{k}{2}}z): 
-\frac{q^{\frac{k}{2}}}{z(1-q^k)}
\right\}\,. 
\label{G- on psipsi*}
\end{eqnarray}

The transformations of $V^{(k)}_m$ 
can be obtained from formulas 
(\ref{G+ on psipsi*}), 
(\ref{G- on psipsi*}) 
by reading the coefficients of the Laurent expansions of 
the equations around $z=0$. 
It is evident but nevertheless surprising 
that $G_{\pm}$ generate automorphisms of 
the Lie algebra (\ref{V-algebra}) by adjoint action.  
Let us concentrate on the transformations of the hamiltonians 
$H_k=V^{(k)}_0$. 
These are read from formulas 
(\ref{G+ on psipsi*}), 
(\ref{G- on psipsi*}) 
as follows. 
\begin{eqnarray}
G_+ V^{(k)}_0 (G_+)^{-1}
&=&
q^{\frac{k}{2}}
\oint 
\frac{dz}{2\pi i} 
\prod_{m=1}^k(1-zq^{\frac{k+1}{2}-m}) 
:\psi(q^{\frac{k}{2}}z)\psi^*(q^{-\frac{k}{2}}z):\,, 
\label{G+ on H_k}
\\
(G_-)^{-1}V^{(k)}_0 G_-
&=&
q^{\frac{k}{2}}
\oint 
\frac{dz}{2\pi i} 
\prod_{m=1}^k(1-z^{-1}q^{-\frac{k+1}{2}+m}) 
:\psi(q^{\frac{k}{2}}z)\psi^*(q^{-\frac{k}{2}}z): \,, 
\label{G- on H_k}
\end{eqnarray}
where the integrals denote taking the residue at $z=0$. 
These integrals can be evaluated by using the $q$-binomial theorem:  
\begin{eqnarray}
\prod_{m=1}^k(1+xq^m)=
\sum_{i=0}^kq^{\frac{i(i+1)}{2}}
\Bigl[
\begin{array}{c}
k \\[-2mm] i 
\end{array}
\Bigr]_q
x^i\,,  
\end{eqnarray}
where 
\begin{eqnarray}
\Bigl[
\begin{array}{c}
k \\[-2mm] i 
\end{array}
\Bigr]_q
=
\frac{(q:q)_k}{(q:q)_i(q:q)_{k-i}}\,, 
\hspace{7mm}
(q:q)_n=(1-q)(1-q^2)\cdots (1-q^n)\,. 
\end{eqnarray} 
By expanding the products in the right hand sides of 
equations (\ref{G+ on H_k}),(\ref{G- on H_k}), 
these are brought to the following form.  
\begin{eqnarray}
G_+ V^{(k)}_0 (G_+)^{-1}
&=&
\sum_{i=0}^k
(-)^iq^{-\frac{i(k-i)}{2}}
\Bigl[
\begin{array}{c}
k \\[-2mm] i 
\end{array}
\Bigr]_q 
V^{(k)}_i\,, 
\label{G+ on H_k 2}
\\
(G_-)^{-1} V^{(k)}_0 G_-
&=&
\sum_{i=0}^k
(-)^iq^{-\frac{i(k-i)}{2}}
\Bigl[
\begin{array}{c}
k \\[-2mm] i 
\end{array}
\Bigr]_q 
V^{(k)}_{-i}\,.  
\label{G- on H_k 2}
\end{eqnarray}

\subsection{Toda equation}
The transformations of $H_1=V^{(1)}_0$ 
in formulas 
(\ref{G+ on H_k 2}),(\ref{G- on H_k 2}) are read as    
\begin{eqnarray}
G_+ H_1 (G_+)^{-1}&=& V^{(1)}_0-V^{(1)}_1\,,  
\label{G+ on H_1}
\\
(G_-)^{-1} H_1 G_-&=& V^{(1)}_0-V^{(1)}_{-1}\,. 
\label{G- on H_1} 
\end{eqnarray} 
These transformations deal naturally 
in the evolution of the partition function 
(\ref{Z_p(t)}) by the time $x=t_1$. 
The operator $G_+e^{H(t)}G_-$ that appears 
in the expression (\ref{Z_p(t) CFT}) evolves according to  
\begin{eqnarray}
\partial_x (G_+e^{H(t)}G_-)
=
G_+H_1(G_+)^{-1}\, 
G_+e^{H(t)}G_- 
=
(V^{(1)}_0-V^{(1)}_1)\, 
G_+e^{H(t)}G_- \,. 
\label{x_evolution of H(t) 1}
\end{eqnarray}
It can be also written as 
\begin{eqnarray}
\partial_x (G_+e^{H(t)}G_-)
=
G_+e^{H(t)}G_- \,\, 
(G_-)^{-1}H_1G_- 
=
G_+e^{H(t)}G_-\,\, 
(V^{(1)}_0-V^{(1)}_{-1})\,. 
\label{x_evolution of H(t) 2}
\end{eqnarray}
These two descriptions lead to 
\begin{eqnarray}
\partial_x Z_p(t)
&=& 
\langle p |\,(V^{(1)}_0-V^{(1)}_1)\, G_+e^{H(t)}G_-\,|p \rangle 
\\
&=& 
\langle p |\,G_+e^{H(t)}G_-\, (V^{(1)}_0-V^{(1)}_{-1})\,|p \rangle\,. 
\end{eqnarray}

Let us derive the Toda equation (\ref{Toda equation}). 
Owing to the identification (\ref{phi_p}),  
it suffices to prove the following identity. 
\begin{eqnarray}
Z_p\, 
\partial^2_xZ_p-(\partial_xZ_p)^2
=
q^{2p+1}Z_{p+1}\,
Z_{p-1}\,, 
\hspace{6mm} 
p \in \mathbb{Z}\,.
\label{Toda equation 2} 
\end{eqnarray}

We first rewrite the left hand side of equation 
(\ref{Toda equation 2}), 
using the expression (\ref{Z_p(t) CFT})
and  
applying formulas (\ref{x_evolution of H(t) 1}), (\ref{x_evolution of H(t) 2}) 
in it,  
as follows. 
\begin{eqnarray}
&&
Z_p\, \partial^2_xZ_p 
-(\partial_xZ_p)^2 
\nonumber \\
&&
=
Z_p \cdot 
\langle p |\,(V^{(1)}_0-V^{(1)}_1)\,
G_+e^{H(t)}G_-\,(V^{(1)}_0-V^{(1)}_{-1})\,|p \rangle 
\nonumber \\
&&
~~
-\langle p |\,(V^{(1)}_0-V^{(1)}_1)\,G_+e^{H(t)}G_-\,|p \rangle 
\cdot 
\langle p |\,G_+e^{H(t)}G_-\,(V^{(1)}_0-V^{(1)}_{-1})\,|p \rangle
\label{D_xD_x 1}
\end{eqnarray}
The matrix elements in the right hand side of this equation 
can be translated to the fermion correlation functions, 
by replacing $V^{(1)}_0-V^{(1)}_{\pm 1}$ 
with the states generated by these operators. 
The corresponding states are
\begin{eqnarray}
(V_0^{(1)}-V^{(1)}_{-1})\,|p\rangle
=q \frac{1-q^p}{1-q}\,|p\rangle 
+q^{p+\frac{1}{2}}\psi_{-p-1}\psi^*_p\,|p \rangle
\end{eqnarray}
and its conjugate state. 
Thereby, 
we can eventually translate 
the right hand side of equation (\ref{D_xD_x 1})   
into the following combination of the correlation functions.   
\begin{eqnarray}
&&
Z_p\, \partial^2_xZ_p-(\partial_xZ_p)^2 
\nonumber \\
&&
=
q^{2p+1}
Z_p 
\cdot 
\langle p |\,\psi_{-p}\psi^*_{p+1} 
              \,\,
           G_+e^{H(t)}G_- 
              \,\, 
            \psi_{-p-1}\psi^*_p \,|p \rangle 
\nonumber \\
&&
~~
-q^{2p+1}
\langle p |\,\psi_{-p}\psi^*_{p+1} 
                \,\, 
           G_+e^{H(t)}G_-\,|p \rangle 
\cdot 
\langle p |\,G_+e^{H(t)}G_-
                \,\, 
           \psi_{-p-1}\psi^*_p\,|p \rangle
\label{D_xD_x 2}
\end{eqnarray}

Wick's theorem shows that 
correlation functions of free fermions 
are factorized into products of 
their two point functions.  
The four point function in the right hand side of 
equation (\ref{D_xD_x 2}) 
is factorized into 
\begin{eqnarray} 
&&
\frac{1}{Z_p} 
\langle p |\,\psi_{-p}\psi^*_{p+1} 
              \,\,
           G_+e^{H(t)}G_- 
              \,\, 
            \psi_{-p-1}\psi^*_p \,|p \rangle 
\nonumber \\
&&
=
\frac{1}{Z_p}
\langle p |\,\psi^*_{p+1} 
              \,\,
           G_+e^{H(t)}G_- 
              \,\, 
            \psi_{-p-1}\,|p \rangle 
\cdot 
\frac{1}{Z_{p}}
\langle p |\,\psi_{-p} 
              \,\,
           G_+e^{H(t)}G_- 
              \,\, 
           \psi^*_p \,|p \rangle 
\nonumber \\
&&
\hspace{10mm}
+
\frac{1}{Z_p}
\langle p |\,\psi_{-p}\psi^*_{p+1} 
              \,\,
           G_+e^{H(t)}G_- \,|p \rangle 
\cdot 
\frac{1}{Z_p}
\langle p |\,G_+e^{H(t)}G_- 
              \,\, 
           \psi_{-p-1}\psi^*_p \,|p \rangle\,.  
\label{Wick theorem}
\end{eqnarray}
In the right hand side of this equation,  
the first term equals, 
making use of the relations 
$|p+1\rangle=\psi_{-p-1}|p \rangle$ 
and 
$|p-1\rangle=\psi^*_{p}|p \rangle$, 
to $Z_p^{-2}Z_{p+1}Z_{p-1}$. 
Thus, 
Wick's theorem leads to 
\begin{eqnarray}
&&
Z_p \cdot
\langle p |\,\psi_{-p}\psi^*_{p+1} 
              \,\,
           G_+e^{H(t)}G_- 
              \,\, 
            \psi_{-p-1}\psi^*_p \,|p \rangle 
\nonumber \\
&&
\hspace{5mm}
=
Z_{p+1}Z_{p-1} 
+
\langle p |\,\psi_{-p}\psi^*_{p+1} 
              \,\,
           G_+e^{H(t)}G_- \,|p \rangle 
\cdot
\langle p |\,G_+e^{H(t)}G_- 
              \,\, 
           \psi_{-p-1}\psi^*_p \,|p \rangle \,. 
\label{Wick theorem 2}
\end{eqnarray}
By plugging this formula into the right hand side of 
equation (\ref{D_xD_x 2}),  
we obtain $q^{2p+1}Z_{p+1}Z_{p-1}$. 
Thereby this completes the proof.


\section{One-dimensional Toda hierarchy}
In this section we prove that 
the series of the partition functions (\ref{tau(t;p)=Z_p}) 
is a tau function of the one-dimensional Toda hierarchy.

We first recall the theory of tau functions of the Toda hierarchy 
\cite{Ueno-Takasaki, Miwa-Jimbo, Takebe}. 
The two-dimensional Toda hierarchy has 
two series of commuting flows and thereby 
two series of time variables,  
$T=(T_1,T_2,\cdots)$ and $\bar{T}=(\bar{T}_1,\bar{T}_2, \cdots)$,  
each of which describes each the commuting flows.  
Tau functions of the two-dimensional Toda hierarchy are admitted 
to have several expressions including the realization 
by means of free fermions or free bosons.  
The standard description in terms of free fermions is 
\begin{eqnarray}
\tau^{2\,Toda}(T,\bar{T};p)=
e^{\sum_{k=1}^{\infty}(c_kT_k+\bar{c}_k\bar{T}_k)}\,
\langle p |\,
e^{\sum_{k=1}^{\infty}T_kJ_k}\,\, g\,\, 
e^{-\sum_{k=1}^{\infty}\bar{T}_kJ_{-k}}\,
| p \rangle\,. 
\label{2Toda tau}
\end{eqnarray}
where 
$g$ is an element of $GL(\infty)$. 
$c_k$ and $\bar{c}_k$ are numerical constants which originate 
in the ambiguity of the tau function. 
The two-dimensional Toda hierarchy 
reduces to the one-dimensional Toda hierarchy   
when the two-sided time evolutions degenerate. 
This reduction imposes 
the following constraint on the tau function. 
\begin{eqnarray}
\Bigl(
\frac{\partial}{\partial T_k}
+
\frac{\partial}{\partial \bar{T}_k}
\Bigr)
\tau^{2\,Toda}(T,\bar{T};p)
=0\,, 
\hspace{8mm}
k=1,2,\cdots\,.
\label{1Toda reduction}
\end{eqnarray} 
When the condition fulfilled, 
the tau function translates to the tau function 
of the one-dimensional Toda hierarchy. 
Owing to the degeneration,  
the one-dimensional Toda hierarchy 
has one series of commuting flows. 
The corresponding time variables can be identified 
with $T=(T_1,T_2,\cdots)$. 
The tau function has the following expression 
in terms of free fermions. 
\begin{eqnarray}
\tau^{1\,Toda}(T;p)=
e^{\sum_{k=1}^{\infty}c_kT_k}\,
\langle p |\,
e^{\frac{1}{2}\sum_{k=1}^{\infty}T_kJ_k}\,\, g\,\, 
e^{\frac{1}{2}\sum_{k=1}^{\infty}T_kJ_{-k}}\,
| p \rangle\,. 
\label{1Toda tau}
\end{eqnarray}
The coupling constants $t=(t_1,t_2,\cdots)$ 
in the series of the partition functions 
(\ref{tau(t;p)=Z_p}) is eventually 
identified with the standard Toda time variables 
$T=(T_1,T_2,\cdots)$ by $T_k=(-)^kt_k$. 

To implement the constraint (\ref{1Toda reduction}), 
$g$ is chosen to satisfy the constraint 
\begin{eqnarray}
J_k\,g\,=\,g\,J_{-k}\,, 
\hspace{8mm}k=1,2,\cdots\,.
\label{1Toda condition of g}
\end{eqnarray}
Under this condition, 
the foregoing expression of $\tau^{1\,Toda}(T;p)$ 
can be rewritten as 
\begin{eqnarray}
\tau^{1\,Toda}(T;p)=
e^{\sum_{k=1}^{\infty}c_kT_k}\,\,
\langle p |\,
e^{\sum_{k=1}^{\infty}T_kJ_k}
\,\,g\, 
|p \rangle\,. 
\label{mKP}
\end{eqnarray}
This shows that $\tau^{1\,Toda}(T;p)$ is a tau function 
of the modified KP hierarchy as well.

\subsection{Shift symmetry}
Among automorphisms of the Lie algebra (\ref{V-algebra}), 
we pay special attention to the following shift symmetry: 
\begin{eqnarray}
V_{m}^{(k)}-\delta_{m,0}\frac{q^k}{1-q^k} 
\,\,\,\,\longmapsto \,\,\,\, 
V_{m+k}^{(k)}-\delta_{m+k,0}\frac{q^k}{1-q^k}\,, 
\hspace{8mm}
k \ge 1\,,  
\label{shift symmetry}
\end{eqnarray}
and $V_m^{(0)}$ are left unchanged. 
Using this symmetry,  
three commutative sub-algebras generated respectively by 
$\bigl\{V^{(k)}_0\bigr\}_{k\geq 0}, 
\bigl\{ V^{(k)}_k\bigr\}_{k\geq 0}$ 
and 
$\bigl\{ V^{(k)}_{-k}\bigr\}_{k\geq 0}$ 
become conjugate to one another.

The symmetry (\ref{shift symmetry}) 
becomes eventually one of the automorphisms  
generated by the adjoint action of $G_{\pm}$. 
Combining the transformations 
(\ref{G+ on psipsi*}),(\ref{G- on psipsi*}),  
we find 
\begin{eqnarray}
&&
(G_-G_+)
\left\{ 
:\psi(q^{\frac{k}{2}}z)\psi^*(q^{-\frac{k}{2}}z): 
-\frac{q^{\frac{k}{2}}}{z(1-q^k)}
\right\}
(G_-G_+)^{-1}
\nonumber \\
&&
\hspace{15mm}
=
(-)^kz^k
\left\{
:\psi(q^{\frac{k}{2}}z)\psi^*(q^{-\frac{k}{2}}z): 
-\frac{q^{\frac{k}{2}}}{z(1-q^k)}
\right\}\,. 
\label{G-G+ on psipsi*}
\end{eqnarray}
Taking account of the mode expansion (\ref{V_m(k)}), 
we can read the symmetry (\ref{shift symmetry}) 
from formula (\ref{G-G+ on psipsi*}) 
in the following form.    
\begin{eqnarray}
(G_-G_+)\,
\left(
V_{m}^{(k)}-\delta_{m,0}\frac{q^k}{1-q^k}
\right)\,
(G_-G_+)^{-1}
=
(-)^k
\left( 
V_{m+k}^{(k)}-\delta_{m+k,0}\frac{q^k}{1-q^k}
\right)\,. 
\label{G-G+ on V_m(k)}
\end{eqnarray}

This formula shows that the conjugacy between 
the three sub-algebras is realized 
as their transformations by $(G_-G_+)^{\pm 1}$. 
We shall describe such transformations in some detail. 
Let $k \geq 1$. Putting $m=0$ in (\ref{G-G+ on V_m(k)}),  
we obtain 
\begin{eqnarray}
(G_-G_+)
\left(V_0^{(k)}-\frac{q^k}{1-q^k}\right) 
(G_-G_+)^{-1}
=
(-)^kV_k^{(k)}\,. 
\label{H_k vs V_k(k)}
\end{eqnarray}
Similarly, putting $m=-k$ in (\ref{G-G+ on V_m(k)}), 
we obtain 
\begin{eqnarray}
(G_-G_+)^{-1}
\left(V_0^{(k)}-\frac{q^k}{1-q^k}\right) 
(G_-G_+)
=
(-)^kV_{-k}^{(k)}\,. 
\label{H_k vs V_-k(k)}
\end{eqnarray}

Let us consider two more commutative sub-algebras 
which are generated respectively by 
$\bigl\{ V_k^{(0)}\bigr\}_{k \ge 0}$ 
and  
$\bigl\{ V_{-k}^{(0)} \bigr\}_{k \ge 0}$. 
These sub-algebras become conjugate to 
the aforementioned three sub-algebras by 
the adjoint actions of $(G_-G_+)^{\pm 1}$ 
and $q^{\pm \frac{W}{2}}$. 
To see this, 
note that W (\ref{W_0(3)}) is able to rotate 
$V^{(k)}_{\pm k}$ to $V_{\pm k}^{(0)}$ 
by the adjoint action.
Actually, 
we have the following transformations.  
\begin{eqnarray}
q^{\frac{W}{2}}\,V^{(k)}_k\,q^{-\frac{W}{2}}
&=&V_{k}^{(0)}\,, 
\label{V_k(k) vs J_k}\\
q^{-\frac{W}{2}}\,V^{(k)}_{-k}\, q^{\frac{W}{2}}
&=&
V_{-k}^{(0)}\,. 
\label{V_-k(k) vs J_-k}
\end{eqnarray}
To obtain these formulas, 
note that, 
by the adjoint action of $q^{\frac{W}{2}}$,  
the fermions transform as   
$q^{\frac{W}{2}}\psi_mq^{-\frac{W}{2}}
=q^{\frac{m^2}{2}}\psi_m$
and 
$q^{\frac{W}{2}}\psi_m^*q^{-\frac{W}{2}}=
q^{-\frac{m^2}{2}}\psi_m^*$. 
By using these transformations, 
the left hand side of (\ref{V_k(k) vs J_k}) 
can be computed as 
\begin{eqnarray}
q^{\frac{W}{2}}\,V^{(k)}_k\,q^{-\frac{W}{2}}
&=& 
q^{-\frac{k^2}{2}}
\sum_{n \in \mathbb{Z}}q^{kn}\,
q^{\frac{W}{2}}:\psi_{k-n}\psi_n^*:q^{-\frac{W}{2}} 
\nonumber \\
&=& 
q^{-\frac{k^2}{2}}
\sum_{n \in \mathbb{Z}}
q^{kn+\frac{(k-n)^2}{2}-\frac{n^2}{2}}
:\psi_{k-n}\psi_n^*:
\nonumber \\
&=&
\sum_{n \in \mathbb{Z}}
:\psi_{k-n}\psi_n^*:\,, 
\end{eqnarray}
which is nothing but $V_k^{(0)}$. 
Thus we obtain the formula (\ref{V_k(k) vs J_k}).
The similar computation leads to the formula (\ref{V_-k(k) vs J_-k}).

The formulas 
(\ref{V_k(k) vs J_k}),
(\ref{V_-k(k) vs J_-k})
in addition to 
(\ref{H_k vs V_k(k)}), 
(\ref{H_k vs V_-k(k)}) 
show that all the five sub-algebras are conjugate 
to one another by the adjoint actions of 
$(G_-G_+)^{\pm 1}$ and $q^{\pm \frac{W}{2}}$. 
Therefore, 
the hamiltonian $H_k=V^{(k)}_0$ can transform to 
$J_{\pm k}=V^{(0)}_{\pm k}$ by the adjoint actions of
$(G_-G_+)^{\pm 1}$ and $q^{\pm \frac{W}{2}}$. 
Actually, 
as seen from formulas 
(\ref{H_k vs V_k(k)}), (\ref{V_k(k) vs J_k}), 
$q^{\frac{W}{2}}(G_-G_+)$ rotates $H_k$ to $J_k$,    
while 
$q^{-\frac{W}{2}}(G_-G_+)^{-1}$ rotates $H_k$ to $J_{-k}$, 
as seen from formulas 
(\ref{H_k vs V_-k(k)}), (\ref{V_-k(k) vs J_-k}).  
We write down these transformations in the following form 
convenient for the later use. 
\begin{eqnarray}
G_+
\left( V_0^{(k)}-\frac{q^k}{1-q^k} \right) 
(G_+)^{-1}
&=&
(-)^k
\bigl( q^{\frac{W}{2}}G_- \bigr)^{-1}
J_k\,\,  
(q^{\frac{W}{2}}G_-)\,,  
\label{V_0(k) vs J_k}\\
(G_-)^{-1}
\left( V_0^{(k)}-\frac{q^k}{1-q^k} \right) 
G_-
&=&
(-)^k
\bigl( G_+q^{\frac{W}{2}} \bigr)\,
J_{-k}\,
(G_+ q^{\frac{W}{2}})^{-1}\,. 
\label{V_0(k) vs J_-k}
\end{eqnarray}

\subsection{The proof}

\subsubsection{New representation of the partition functions}

We derive the expression (\ref{Z_p(t) Toda}) 
of the partition function (\ref{Z_p(t)}). 

Let us first rewrite $G_+e^{H(t)}G_-$ as follows. 
\begin{eqnarray}
G_+e^{H(t)}G_- 
&=& 
G_+e^{\frac{1}{2}H(t)}\,e^{\frac{1}{2}H(t)}G_- 
\nonumber \\
&=& 
G_+e^{\frac{1}{2}H(t)}(G_+)^{-1} 
G_+G_-\,\, 
(G_-)^{-1}
e^{\frac{1}{2}H(t)}G_- \,. 
\label{factorization 1}
\end{eqnarray}
The transformations of $e^{\frac{1}{2}H(t)}$ 
by the adjoint actions of $G_+$,$(G_-)^{-1}$  
in this expression can be evaluated 
by using formulas 
(\ref{V_0(k) vs J_k}), (\ref{V_0(k) vs J_-k}). 
Eventually, these transformations are expressed as 
\begin{eqnarray}
G_+\, 
e^{\frac{1}{2}H(t)}\,
(G_+)^{-1} 
&=&
e^{\sum_{k=1}^{\infty}\frac{t_kq^k}{2(1-q^k)}}\,\,
(q^{\frac{W}{2}}G_-)^{-1}\, 
e^{\frac{1}{2}\sum_{k=1}^{\infty}(-)^kt_kJ_k}\,\, 
(q^{\frac{W}{2}}G_-) 
\label{transformation 1}
\\
(G_-)^{-1}\, 
e^{\frac{1}{2}H(t)}\, 
G_- 
&=& 
e^{\sum_{k=1}^{\infty}\frac{t_kq^k}{2(1-q^k)}}\,\,
(G_+q^{\frac{W}{2}})\, 
e^{ \frac{1}{2}\sum_{k=1}^{\infty}(-)^kt_kJ_{-k}}\,\, 
(G_+q^{\frac{W}{2}})^{-1}
\label{transformation 2}
\end{eqnarray}
By plugging these expressions 
into the right hand side of (\ref{factorization 1}),  
we obtain the following formula. 
\begin{eqnarray}
G_+e^{H(t)}G_- 
&=& 
e^{\sum_{k=1}^{\infty}\frac{t_kq^k}{1-q^k}}\,\,
(q^{\frac{W}{2}}G_-)^{-1}\,\, 
e^{\frac{1}{2}\sum_{k=1}^{\infty}(-)^kt_kJ_k}\,\,
\nonumber  \\
&&
\times\,
{\bf g}_{\star}
\,\, 
e^{\frac{1}{2}\sum_{k=1}^{\infty}(-)^kt_kJ_{-k}}\,\, 
(G_+q^{\frac{W}{2}})^{-1}\,, 
\label{factorization 2}
\end{eqnarray}
where ${\bf g}_{\star}$ is the element of $GL(\infty)$ 
given by (\ref{g_star}).

Making use of this formula we arrange the expression 
(\ref{Z_p(t) CFT}) as 
\begin{eqnarray} 
&&
Z_{p}(t)=\langle p| G_+e^{H(t)}G_- | p \rangle 
\nonumber \\
&&
= 
e^{\sum_{k=1}^{\infty}\frac{t_kq^k}{1-q^k}}\,\,
\langle p |\,
(G_-)^{-1}\,
q^{-\frac{W}{2}}\,\, 
e^{\frac{1}{2}\sum_{k=1}^{\infty}(-)^kt_kJ_k}
\,\,
{\bf g}_{\star}
\,\, 
e^{\frac{1}{2}\sum_{k=1}^{\infty}(-)^kt_kJ_{-k}}\,\, 
q^{-\frac{W}{2}}\,
(G_+)^{-1}\,
|p \rangle
\nonumber \\
&&
= 
e^{\sum_{k=1}^{\infty}\frac{t_kq^k}{1-q^k}}\,\,
\langle p |\,
q^{-\frac{W}{2}}\,\, 
e^{\frac{1}{2}\sum_{k=1}^{\infty}(-)^kt_kJ_k}
\,\,
{\bf g}_{\star}
\,\, 
e^{\frac{1}{2}\sum_{k=1}^{\infty}(-)^kt_kJ_{-k}}\,\, 
q^{-\frac{W}{2}}\,
|p \rangle\,. 
\label{Z vs Toda 1}
\end{eqnarray}
Considering in the last line  
that the $W$-charge of the state $|p\rangle$ is 
$\frac{1}{6}p(p+1)(2p+1)$, 
we finally obtain the expression (\ref{Z_p(t) Toda}).

\subsubsection{Reduction to one-dimensional Toda hierarchy}

Let us show that 
$\tau(t;p)=q^{\frac{1}{6}p(p+1)(2p+1)}Z_{p}(t)$ is 
a tau function of the one-dimensional Toda hierarchy. 
The key is the fact that ${\bf g}_{\star}$ 
satisfies (\ref{1Toda condition of g}). 
This can be seen by using formulas (\ref{H_k vs V_k(k)}), 
(\ref{H_k vs V_-k(k)}), (\ref{V_k(k) vs J_k}), (\ref{V_-k(k) vs J_-k}) 
as follows. 
\begin{eqnarray}
J_k\,{\bf g}_{\star}
&=&
J_kq^{\frac{W}{2}}(G_-G_+)^2q^{\frac{W}{2}} 
\nonumber \\
&=& 
q^{\frac{W}{2}}V_k^{(k)}(G_-G_+)^2q^{\frac{W}{2}} 
\nonumber \\
&=& 
q^{\frac{W}{2}}(G_-G_+)\, 
(-)^k\Bigl(V_0^{(k)}-\frac{q^k}{1-q^k}\Bigl)\, 
(G_-G_+)q^{\frac{W}{2}}
\nonumber \\
&=& 
q^{\frac{W}{2}}(G_-G_+)^2 
V^{(k)}_{-k}q^{\frac{W}{2}}
\nonumber \\
&=& 
q^{\frac{W}{2}}(G_-G_+)^2q^{\frac{W}{2}} 
J_{-k}
\nonumber \\
&=&
{\bf g}_{\star}J_{-k}\,. 
\end{eqnarray}
The constraint (\ref{1Toda condition of g}) 
is equivalent to the constraint (\ref{1Toda reduction}) 
on the tau function given by (\ref{2Toda tau}) 
taking $g={\bf g}_{\star}$. 
This means that ${\bf g}_{\star}$ actually gives a solution of 
the one-dimensional Toda hierarchy. 
Therefore it follows from (\ref{1Toda tau}) that 
$\tau(t;p)$ is the corresponding tau function 
by the identifications $T_k=(-)^kt_k$. 
This completes the proof.


\section{Conclusion and discussion}

We investigated melting crystal, 
which is known as random plane partition, 
from the viewpoint of integrable systems. 
We proved that 
a series of partition functions of 
melting crystals gives rise to a tau function of 
the one-dimensional Toda hierarchy, 
where the models are defined by 
adding suitable potentials, 
endowed with a series of coupling constants, 
to the standard statistical weight. 
We showed that these potentials are converted to 
a commutative sub-algebra of quantum torus Lie algebra.  
Further exploiting the underlying algebraic structure,  
a remarkable connection  between random plane partition 
and quantum torus Lie algebra was revealed. 
This connection substantially enabled to prove 
the statement. 
Based on the result, 
we briefly argued the integrable structures of 
five-dimensional $\mathcal{N}=1$ supersymmetric gauge theories 
and 
$A$-model topological strings. 
The aforementioned potentials correspond to 
gauge theory observables analogous to the Wilson loops, 
and thereby the partition functions are translated 
in the gauge theory 
to generating functions of their correlators. 
In topological strings, 
we particularly comment on a possibility of topology change 
caused by condensation of these observables, 
giving a simple example.

In four-dimensional $\mathcal{N}=2$ supersymmetric gauge theories, 
the authors of \cite{Nekrasov-Marshakov} 
obtained the generating function of correlation functions 
of the higher Casimir operators in the fermionic form 
\begin{eqnarray}
Z_p^{4d U(1)}(x)\,=\,
\langle p |
e^{\frac{1}{\hbar}J_1}\, 
e^{\sum_{k=0}^{\infty}\frac{x_k}{(k+1)!}\mathcal{P}_{k+1}}\,
e^{\frac{1}{\hbar}J_{-1}}
| p \rangle\,, 
\label{Z_p_4dU(1)}
\end{eqnarray}
where $\mathcal{P}_{k}$ are the fermion bilinear forms 
introduced by Okounkov and Pandharipande \cite{OK1}, 
and $x_k$ are the coupling constants of 
the higher Casimirs of the gauge theory.
The above partition function also appears 
in the Gromov-Witten theory as 
the generating function of 
the absolute Gromov-Witten invariants on 
$\mathbb{C}P^1$ \cite{OK1, Nekrasov-Losev}. 
Before this fermionic representation was presented, 
Getzler had conjectured, and later proven \cite{Getzler}, 
that the generating function is a tau function of 
the one-dimensional Toda hierarchy.  
Getzler's proof is, however, fairly complicated and somewhat indirect, 
combining the Virasoro conjecture \cite{Givental} 
with the partial result that the Toda equation holds 
on the subspace $x_2 = x_3 = \cdots = 0$.  
It will be therefore an interesting problem 
to give a more direct proof on the basis of the fermionic representation.
A possible scenario will be, as we have done in the five-dimensional case, 
to find a suitable analogue of ${\bf g}_{\star}$ and 
to rewrite the foregoing fermionic representation 
into a standard form like (\ref{1Toda tau}) and (\ref{mKP}). 
Unfortunately, as we remarked in the end of section 1.5.2, 
the naive four-dimensional ($R \to 0$) limit 
of ${\bf g}_{\star}$ itself does not work. 
Since the role of ${\bf g}_{\star}$ reminds us of 
various ``dressing operators'' 
in the work of Okounkov and Pandharipande \cite{OK1, OK2}, 
a correct four-dimensional analogue of ${\bf g}_{\star}$ 
might be hidden therein.


\end{document}